# Cytoskeleton-inspired, adaptive nanolipogels as superlubricating delivery vehicles

*Panpan Zhao*[1], Avijit Mondal[1], Nir Kampf[1], Aleksei Solomonov[1], Roman Kamyshinsky[2], Jacob Klein*[1]*

1. Department of Molecular Chemistry and Materials Sciences, Weizmann Institute of Science, Rehovot, 76100, Israel.
2. Department of Chemical Research Support, Weizmann Institute of Science, Rehovot, 76100, Israel.

Emails: panpan.zhao@weizmann.ac.il, jacob.klein@weizmann.ac.il

**Abstract**:

Phosphatidylcholine liposomes fill a special niche in alleviating osteoarthritis via intra-articular (IA) administration, attributed to their superlubricity at the articular cartilage surface, but their co-utilization as drug delivery vesicles in such therapy remains challenging as they may rupture under mechanical stress. Here, we describe cytoskeleton-inspired, supramolecular, self-assembled nanolipogels (NLGs), encompassing liposome-encased nanogels with a dynamic network formed by hydrogen bonding and cation-π interactions, as a platform for simultaneous robust drug-delivery and massive reduction of interfacial frictional dissipation. We use a surface force balance to assess such dissipation at the sub-nanometer level, elucidating the mechanism involved, and atomic force microscopy to probe the NLGs structural stability. A useful proxy for the interfacial dissipation is the coefficient of friction, which remains as low as $\sim 10^{-4}$ at contact pressures at least up to 2 MPa, while under higher pressures exceeding the H-bonding energy density it increases abruptly and irreversibly to the still-low value $\sim 10^{-2}$. Under sustained sliding above this threshold, however, friction gradually decreases again, indicating recovery of the lubricating interface. Molecular dynamics simulations identify the compressive stress decrease due to hydrogen-bond rupture/rearrangement within the nanogel as a buried supramolecular transition associated with lubrication breakdown and recovery, while cargo release during sliding emphasizes the drug-delivery potential of such NLGs. These findings reveal how supramolecular core-shell reinforcement regulates load-bearing hydration lubrication, and provides a framework for designing adaptive biomimetic lubricants which are at the same time load-bearing intra-articular cargo-delivery vehicles.

ToC graphic:

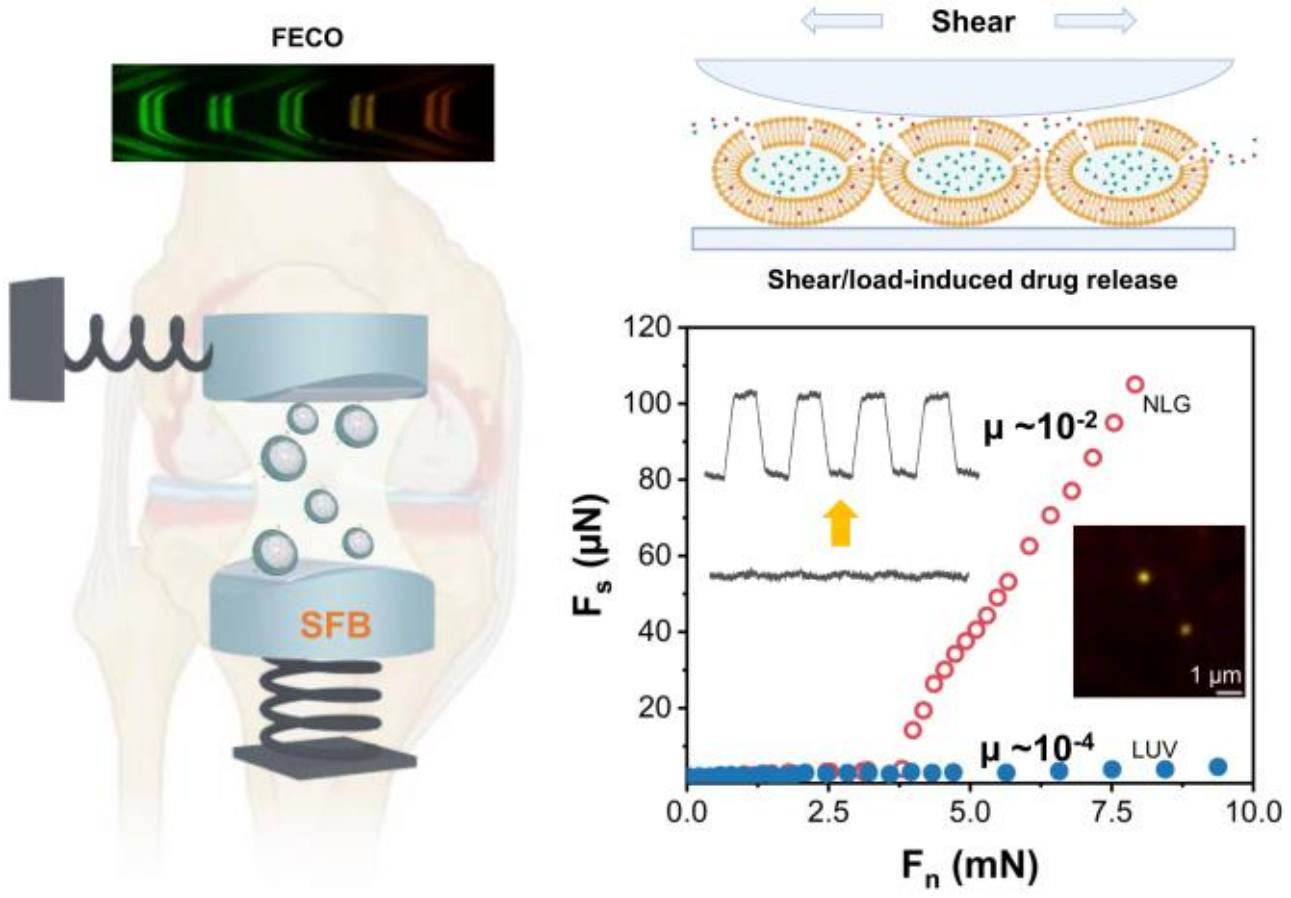

FECO
SFB
Shear
Shear/load-induced drug release
μ ~10⁻²
NLG
μ ~10⁻⁴
LUV
1 μm
Fs (μN)
Fn (mN)

## Introduction

Osteoarthritis (OA), a painful, debilitating joint disease affecting hundreds of millions worldwide[1] may be viewed as a pathology associated with excessive interfacial dissipation and wear (resulting from lubrication breakdown) on articulation of cartilage, characterized by inflammation and progressive tissue degeneration.[2, 3] For this reason, two important aims for treatment for OA are, on the one hand, to control inflammation using intra-articular drug administration, and at the same time to reduce interfacial dissipation and resulting wear damage to the rubbing cartilage surfaces, while still maintaining the function of the articular joint.[4] Phosphatidylcholine (PC) vesicles (liposomes) with highly-hydrated PC headgroups exposed at their outer surface, are promising candidates both for drug delivery[5] and for providing a massive reduction in interfacial sliding dissipation[6, 7]. A useful proxy for the latter is the coefficient of friction (COF) μ = [force to slide/load], as follows. The interfacial energy dissipation ΔE between two flat surfaces in contact via an interfacial layer, sliding past each other in the *x*-direction, over an area A($x$), under a mean contact pressure P($x$), is given by

$$\Delta E = \int^{x} \mu(x, dx/dt).\mathrm{A}(x).\mathrm{P}(x).\mathrm{d}x, \qquad (1)$$

where μ is the friction coefficient at $x$ and under sliding velocity (dx/dt) = $v_s$. Since the other parameters are either geometrical (A) or external constraints (P), μ thus serves as a proxy for quantification of the interfacial energy dissipation, reflecting the properties of the sheared interfacial layer itself. PC liposomes have demonstrated $\mu \approx 10^{-3}$ to $10^{-4}$ up to the physiologically highest contact pressures of 10 MPa or more, which has been measured at cartilage surfaces[8], and have garnered attention for their potential to alleviate and locally treat lubrication-associated OA via intra-articular (IA) injection.[9, 10]. Indeed, recent clinical trials have shown significant alleviation of OA pain upon intra-articular injection of lipid-based lubricants[10, 11]. Challenges remain, however, regarding their physical and chemical stability to fusion or aggregation when liposomes are applied as drug delivery vehicles.[10, 12] PEGylation (i.e. functionalizing the liposomes with PEG, poly(ethylene glycol)) is commonly used to stabilize liposomes via steric stabilization.[13] However, clinical applications of PEGylated liposomal vehicles encapsulating hydrophilic drugs in their inner space, may fail due to rupturing, and subsequent leakage and loss of the drugs, when adsorbed on negatively charged surfaces[14], including cartilage,[15] as well as most biological surfaces[16]. Thus liposome-based nanocarriers that retain structural stability on negatively-charged surfaces, and at the same time provide the outstanding reduction in interfacial dissipation associated with lipid bilayers, hold strong potential for OA treatment via their intra-articular injection (as recently demonstrated[11]); their design (and elucidation of their properties) is the challenge addressed in the present work.

Nanolipogels (NLGs) are core-shell assemblages, where a lipid bilayer encapsulates and is attached to and supported by a gelled core comprising a physically or chemically crosslinked, hydrophilic, macromolecular network, and show promise for applications ranging from drug delivery to tissue engineering.[17] NLGs have a loose conceptual resemblance to the cytoskeleton, which is a highly dynamic and adaptive network of filamentous polymers that exists in the cytoplasm of cells, providing mechanical support to cell membranes and enabling their functions by transmitting stress over cellular length scales.[18] Although this supramolecular core-shell reinforcement analogy is clearly limited, as the living cytoskeleton is a dynamic, adaptive protein network capable of active remodeling and self-repair, some self-repair aspects of this cell-inspired concept do apply to our NLGs, as we shall later see. The flexible hydrogel core thus not only provides an inherent cushion against the rupture of the lipid bilayers attached to it, by distributing and thus reducing local stresses, but, in particular, may also serve as a cargo space for the encapsulation and controlled release of a wide variety of drugs.[19]. At the same time, the attached, supported phospholipid bilayer membrane is more stable than other PC lipid vectors, such as liposomes, while preserving bilayer fluidity.[20] Previous investigations of NLGs mainly focused on their stable encapsulation, cell/tissue targeting and controlled drug release[21, 22]; however, exploiting such NLGs for modulating interfacial dissipation properties has not to our knowledge been attempted.

With the cytoskeleton in mind, we therefore designed an NLG combining features of both the liposome and the soft nanogel: an extreme reduction of dissipation on interfacial sliding, achieved via the highly-hydrated PC headgroups; together with dynamic hydrogen-bonding networks to mimic the plasticity of the cytoskeleton. Structural examination with atomic force microscopy (AFM) show clearly that PEGylated liposomes rupture when adsorbed onto negatively charged (mica) surfaces, whereas NLGs maintained their structural integrity due to the support and stress-distribution provided by the enclosed dynamic gel networks. In particular, we used a surface force balance (SFB) with extreme sensitivity in measuring both normal and lateral surface interactions to systematically investigate the ability of NLGs to modulate dissipation between sliding negatively-charged surfaces to which they are attached. The compressive stress–strain behavior of the interior network, indicated by molecular dynamics simulations, revealed a supramolecular transition underlying lubrication failure and subsequent recovery. This sheds strong light on the nature of the compressed surface boundary layers at the nanometer level, as well as changes that may occur upon compression and sliding. Our findings point to design principles for intra-articularly-administered nanolubricants: if the interior supramolecular network is engineered correctly, one may combine low friction, mechanical resilience, and drug-delivery function in a single platform.

## Results and Discussion

### Synthesis and structural characterization

Dynamic physically cross-linked networks (e.g., hydrogen-bonded networks) enable easy processability, tunable strength and stability, and malleability of the hydrogels, in contrast to covalently cross-linked networks[23]. Loosely inspired by the cytoskeleton with its dynamic networks attached to the encapsulating cell membrane (Figure 1A), a nanosized hydrogel (PTN) was prepared from polyvinyl alcohol (PVA) and tannic acid (TA) via hydrogen-bonding interactions. Here, the stronger hydrogen bonds between PVA and TA serve as the "permanent" crosslinkers (are more stable, helping maintain the structural integrity of the hydrogel over time), and the weaker hydrogen bonds within PVA domains act as the "temporary" crosslinkers (can break and reform more easily, contributing to the hydrogel's flexibility and self-healing ability).[21, 24] This nanosized core was then in turn incorporated within a lipid bilayer to form the nanolipogel (NLG) according to the pathway schematically illustrated in Figure 1B. Initially, a lipid film that consists of saturated hydrogenated soybean phosphatidylcholine (HSPC)/PEG-conjugated distearylphosphorylethanolamine (DSPE-PEG, 1.5 mol%) was formed by conventional solvent evaporation process. This then directly self-assembles on the surface of PTN core via driving forces of cation-π and hydrogen bonding interactions between headgroups of phospholipids and aromatic rings/hydroxyls exposed at surfaces of nanogels. Regular liposomes in the form of large unilamellar vesicles (LUV) were also constructed via the film hydration-extrusion technique, as previously described,[14] for use as controls.

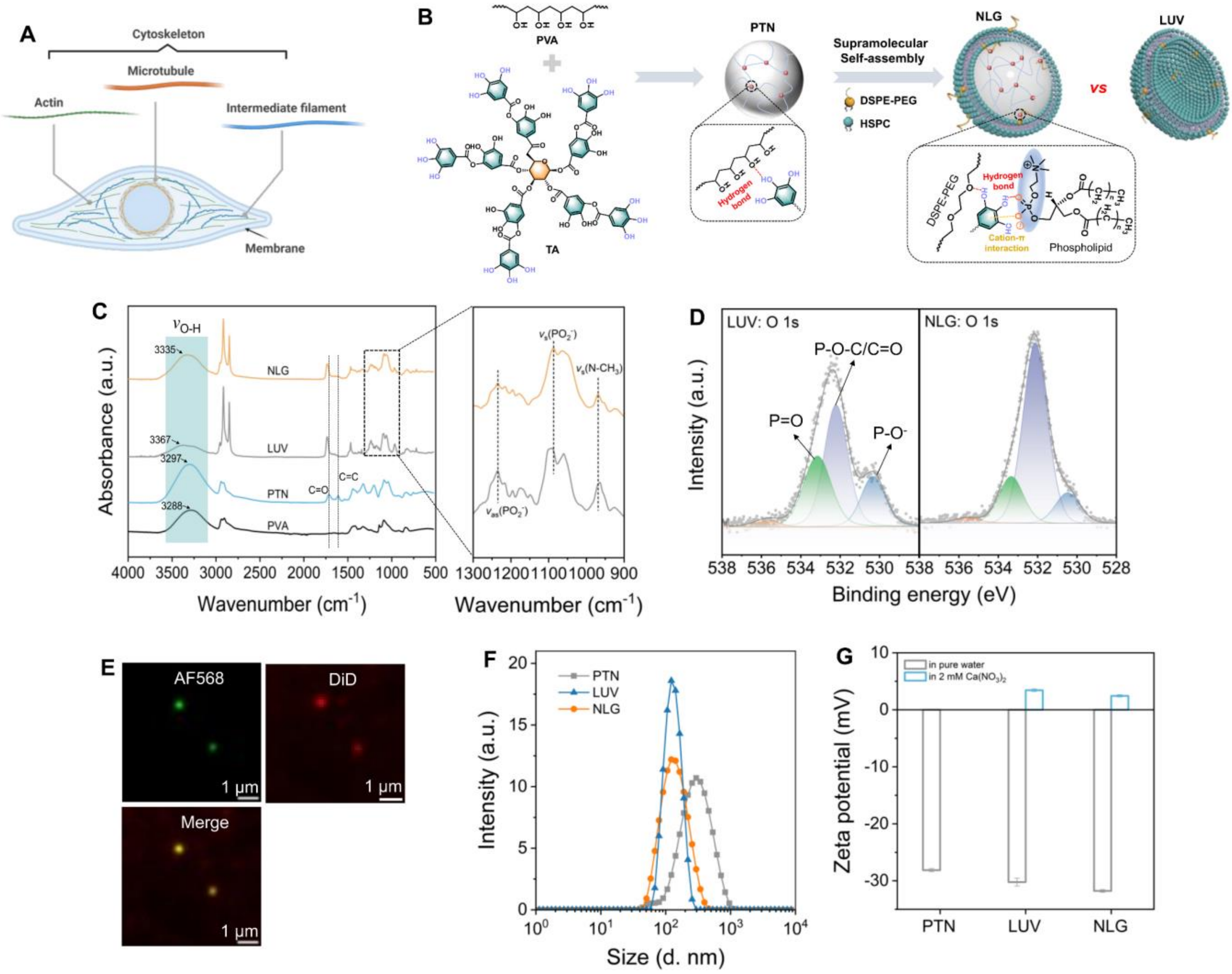

**Figure 1**. (A) Schematic representation of the cytoskeleton. (B) Schematic illustration of supramolecular self-assembly of NLG. (C) ATR-FTIR spectra of PVA, PTN, LUV and NLG. (D) High-resolution XPS O 1s spectra of LUV and NLG were used to analyze lipid-nanogel interactions. (E) Representative confocal fluorescence micrographs of NLG. Interior PTN gels are labelled with dye AF568 (green); lipid membranes are labelled with dye DiD (red); A merge of those two labeling (yellow). (F) DLS diagram of PTN, LUV and NLG in pure water. (G) Zeta potential of PTN, LUV and NLG in pure water and in 2 mM $Ca(NO_3)_2$ salt solution.

The components of PTN and NLG, and the interactions—including cation-π and hydrogen bonding interactions—between lipids and the PTN core were then confirmed by the Attenuated Total Reflection Fourier Transform Infrared Spectroscopy (ATR-FTIR) (Figure 1C). The adsorption peaks around 1713 $cm^{-1}$ (C=O) and 1607 $cm^{-1}$ (C=C) in the spectra of PTN are the specific peaks of ester bonds and benzene rings in TA. NLG shows the typical lipidic backbone bands: $\delta_s(CH_2)$ at 1467 $cm^{-1}$; $\nu_s(CH_2)$ at 2849 $cm^{-1}$; $\nu_{as}(CH_2)$ at 2916 $cm^{-1}$,[25] as well as the typical bonds from PTN: $\nu$(O-H) at 3335 $cm^{-1}$. Those locations of the methylene symmetric ($\nu_s$ ) and asymmetric ($\nu_{as}$ ) stretching of alkyl chains are diagnostic markers of chain conformational order of lipid bilayers in solid crystalline phases.[26] Other bands are phosphate groups ($PO_2^-$) in the headgroup region of NLG: $\nu_s$(1125-1024 cm$^{-1}$) and $\nu_{as}$(1235 cm$^{-1}$) vibrations,[27] but appear to differ from the LUV spectra. The peak for $\nu_s(PO_2^-)$ shifts to a lower wavenumber, while the C-N symmetric stretching mode of the $N^+(CH_3)_3$ group in NLG is found at 970 cm$^{-1}$,[28] which is slightly higher than the corresponding mode in LUV (963 cm$^{-1}$). These changes clearly demonstrate an existing interaction between the headgroups of the lipids with embedded nanogels. The downshift of the ν(P–O) band indicates strengthened hydrogen bonding between $PO_2^-$ and OH groups exposed on the nanogel surface, which restricts the vibrational degrees of freedom of the P–O bonds and thereby decreases their resonant frequency. The upshift of the C–N stretching band of the choline headgroup signifies a more constrained electrostatic environment around the quaternary ammonium, consistent with cation–π interactions between $N^+(CH_3)_3$ and the aromatic rings, which stiffen the C–N bonds and result in an increased vibrational frequency.[29] In the XPS O 1s spectra (Figure 1D), the pronounced increase in the P-O-C/C=O component and the concomitant decrease in the P-O$^-$ component for NLG compared with LUV indicate the presence of hydrogen-bonding interactions between OH groups of the nanogel and head groups of the lipid bilayer. Confocal microscopy, together with the overlay of the fluorescence images obtained, Figure 1E, confirms the co-existence of nanogels and lipid bilayers.

The hydrodynamic radii of LUV, PTN and NLG were determined by DLS (Figure 1F). The peak dimension of the NLG is 151.8 ± 3.1 nm, which is smaller than PTN (332.1 ± 6.3 nm), likely due to the filter effect during the extrusion processes,[21] while LUV has the peak hydrodynamic diameter of 134.3 ± 2.1 nm. The polydispersity indexes of 0.197 for NLG and 0.087 for LUV (Table S1) indicate a more uniform size distribution of the latter (in line with earlier studies), though the range of sizes for the NLG is quite large, and, in particular, extends to some 300 – 400 nm for the largest particles. This is relevant for later understanding of the surface interaction profiles with NLGs. As seen in Figure 1G, LUVs, which include DSPE-PEG (negatively-charged arising from phosphate groups), have a large negative zeta (ζ) potential (-30.2 mV), though comparable with literature values[30]. The ζ-potential for NLG (-31.7 mV) is similar to that of LUV, but slightly lower than the charge of PTN (ζ-potential = -28.1 mV). At the 2 mM $Ca(NO_3)_2$ concentration, which is roughly the physiological-level concentration of divalent ions in synovial fluid[31], NLG and LUV form clear dispersions (Figure S1), indicating they do not aggregate and precipitate, and have low positive surface potential: ζ(NLG) =

2.4 ± 0.2 mV and ζ(LUV) = 3.4 ± 0.3 mV, the charge reversal being attributed to adsorbed $Ca^{2+}$ ions. The zeta potential of PTN in 2 mM $Ca(NO_3)_2$ was not measurable as the nanogel precipitates (Figure S1), attributed to cations-induced bridging of polymer chains. TGA measurement (Figure S2) shows that the water content of PTN is about 75 wt%. We note that despite their large negative zeta potential, both LUVs and NLGs (as we see later) adsorb onto negatively-charged mica from the 2 mM $Ca^{++}$ solution, which we attribute to bridging by the positive divalent ions. Since at the higher salt concentrations of physiological fluid (ca. 0.15 M $Na^{+}$ solution) the zeta potentials are much lower (SI, Table S2), we expect adsorption of the nanoparticles (LUV or NLG) onto negatively-charged surfaces to be even more readily mediated by the divalent ions present at similar concentrations (ca. 2mM) in the physiological saline.

**The fluidity of lipid bilayers**

Fluidity generally refers to the high molecular mobility within the lipid bilayer, which is essential for reorientation or reconstruction of the bilayer plane by allowing lipids to move laterally under shear stress during rearrangement of the H-bond network. We conducted differential scanning calorimetry (DSC) to monitor the transition temperature ($T_m$) of the lipid bilayers from a rigid ordered gel phase to a disordered fluid phase.[32] The calorimetric results (Figure 2A) show that LUV dispersion experienced two endothermic peaks, a minor one at 49.8 °C (peak 1) and the main peak at 53.6 °C (peak 2) corresponding, as previously reported, to ordered-to-rippled gel pretransition and rippled gel-to-liquid crystalline transition of lipid bilayers with the melting of acyl chains of the HSPC lipids, respectively.[33] In contrast, those two peaks in the NLG thermogram were much-reduced and merged to a broad peak that was located at close to $T_m$ (53 °C). This directly indicates that the lipid-headgroup/nanogel interactions lead to disorder, and thus a more fluid behaviour of the lipid bilayer structure. This manifests as a much weaker signal upon the phase transition, as seen in Figure 2A for the NLG relative to the unperturbed solid-ordered lipid phase of the LUV.

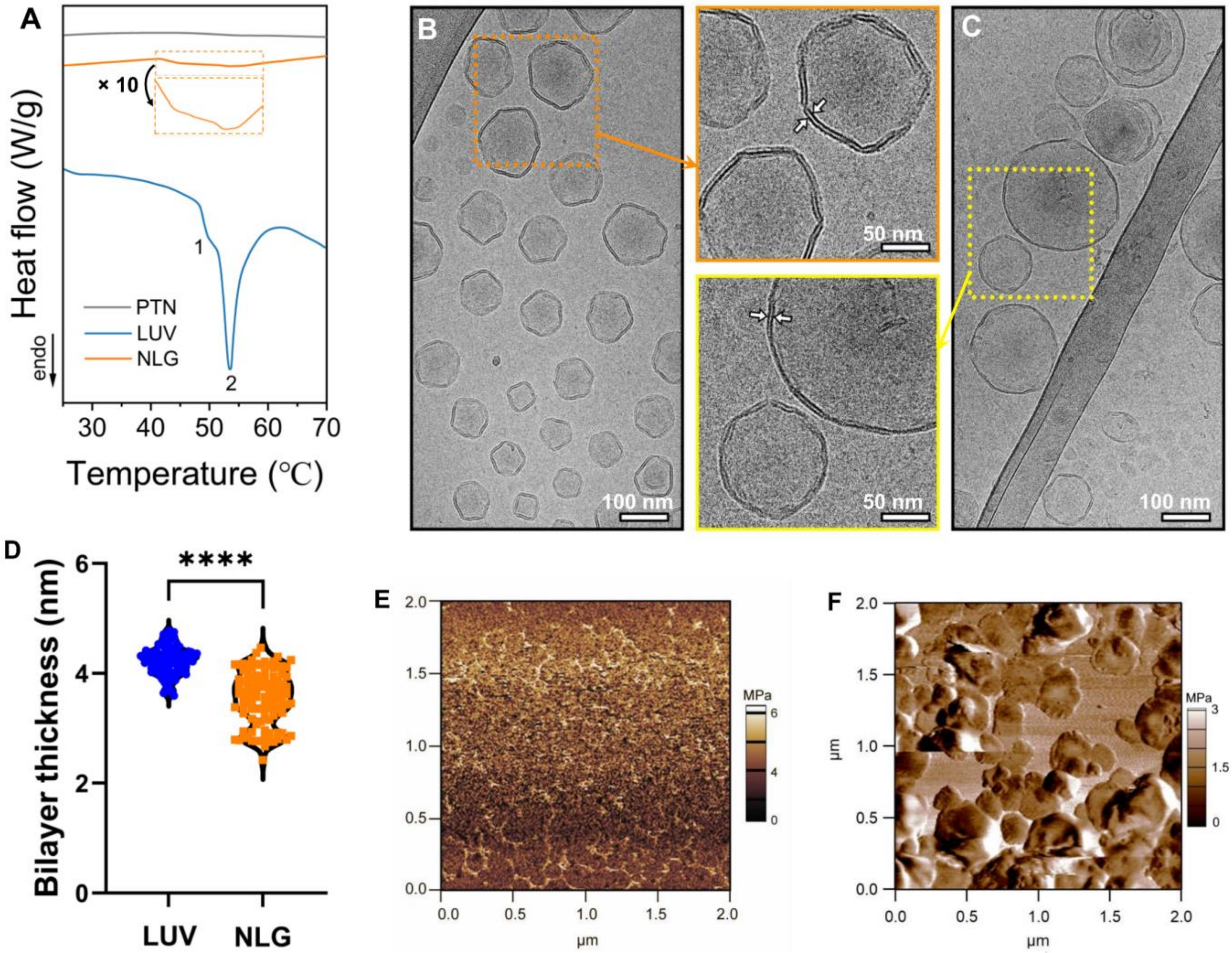

**Figure 2**. The fluidity of lipid bilayer attached to/supported by the nanogel. (A) DSC curves of PTN, LUV and NLG in pure water. Cryo-TEM images of (B) LUV and (C) NLG. Enlarged images (middle) of the areas in the boxes in B and C. Note that the interior nanogel of NLG is invisible to the cryo-TEM due to insufficient contrast. The contrast between the exterior and interior of the bilayers is more or less the same for LUVs and NLGs of the similar sizes, which may be attributed to the low polymer density within the nanogel. (D) Quantification of the thickness of lipid bilayers from Cryo-TEM, counted by Image J (130 points for each sample, see details in SI, Figure S3 and section 4). Statistical significance: **** $p < 0.0001$. AFM mapping of Young's modulus of the mica surfaces bearing LUV (E) and NLG (F).

Cryo-TEM was used to evaluate bilayer thickness in LUVs and NLGs. It is particularly sensitive to the electron-dense phosphates within the phospholipid bilayer, which are visualized as two dark lines bordered by brighter bands in the image.[34] The thickness of the bilayer should, therefore, approximately correspond to the distance between these two dark lines. The quantitative analysis of bilayer thickness measured as 3.58 ± 0.61 nm for NLGs (some of which inevitably have a multilayer structure) versus 4.23 ± 0.26 nm for LUVs (Figure 2D). The difference between them is significant, and is strongly consistent with the increased fluidity of the NLG bilayers relative to the LUV bilayers, indicated by the DSC (Figure 2A) results, since fluid-phase lipid bilayers, for a given tail length, are known to be thinner than gel-phase bilayers. In addition, the cryo-TEM images of LUVs demonstrate homogeneous particles with faceted morphology (Figure 2B), in line with earlier studies of gel-phase liposomes[35], while NLGs appear rounder (Figure 2C), consistent with the more fluid bilayer indicated by the DSC results (Figure 2A). AFM Young's modulus mapping (Figure 2E, F) also revealed directly that the NLG was softer than the lipid bilayer of LUV, as expected since compression of the NLG occurs more easily due to squeeze-out of water from the enclosed nanogel.

**Interfacial Dissipation Determination**

We used a surface forces balance (SFB), the main features of which are schematically outlined in Figure 3A, to directly measure normal forces ($F_n$), and shear forces ($F_s$), (both with a resolution of ca. 100 nN) between two opposing surfaces at absolute separations $D$ (with sub-nanometer resolution). Freshly cleaved mica surfaces were preincubated with NLGs in 2 mM $Ca(NO_3)_2$ solutions for 2 h, and the surface morphology of adsorbed species was then characterized by tapping-mode atomic force microscopy (AFM). The stably adsorbed NLGs on mica substrate indicate an intact nanoparticle configuration with local thickness variations of ~80 nm (Figure 3B), though the absolute thickness of the NLGs on the mica substrate cannot be determined using AFM (it can however be measured using the SFB, see below). As the NLG-bearing mica surfaces across their dispersions are progressively compressed, the frictional forces $F_s$ between them, corresponding to the interfacial sliding dissipation, are measured at different loads $F_n$ and corresponding separations $D$ as the surfaces slide past each other. Typical shear-force $F_s(t)$ vs. time ($t$) traces, recorded directly from the SFB, are shown in Figure 3C. The top triangular waveform shows the back-and-forth lateral motion $\Delta x_0$ of the upper surface as a function of time, while the lower traces correspond to shear (frictional) forces transmitted to the lower surface at the different loads (showing the corresponding contact pressures). The sliding friction for the NLGs corresponds to the flat plateau in the shear-force $F_s(t)$ traces in Figure 3C, and their variation with load is summarized in Figure 3D. The interfacial dissipation behaviour between sliding surfaces bearing NLGs is load-dependent, with two distinct regimes, as shown in Figure 3D. On first approaches between the NLG-bearing surfaces, and at loads corresponding to low and intermediate contact pressures up to ca. 2.2 - 2.5 MPa (comparable with mean cartilage contact pressures in human synovial joints[36, 37]), the interfacial dissipation is very low, with a corresponding CoF $\mu \approx 1.8 \times 10^{-4}$. This is much lower than in reported studies of nanogels-alone/nanoparticles ($\mu$ > 0.03) where the dissipation is attributed to a "rolling" mechanism,[7, 38] but rivals that observed in the LUV system (fig. 4D), as expected since slip occurs at the (presumably) unperturbed bilayers exposed by the NLGs. In the second regime, at high contact stresses (> ca. 2.2 MPa) the dissipation abruptly increases, with a CoF approaching $\mu \approx 10^{-2}$ at the highest loads; we note that even these higher CoF values are well within the range of reported friction coefficients for articular cartilage[39]. Crucially, however, upon repeated sliding at a pressure slightly above this threshold, the friction force decreased by 7-fold after ~600 cycles (blue circles in Figure 3E), recovering the previous low CoF value; the corresponding shear traces are shown in Figure 3F. This indicates that following initial disruption, the interface could recover its low-dissipation configuration. Upon further compression, however, the reconstruction of a continuous lipid-bilayer slip plane becomes difficult, and the cycle-dependent variation in $F_s$ becomes less pronounced (black squares in Figure 3E). On a second approach of the surfaces at a specific contact point, the shear forces were consistently similarly higher ($\mu \sim 10^{-2}$), already starting from low pressures, as expected from non-continuous lipid-bilayers at the contact region. However, the shear force decreased gradually over successive sliding cycles (Figure S4), indicating the lubrication might be recovering slowly. Sliding velocity $v_s$ dependence of the frictional force (Figure S5), which shows only a very weak variation (a 20-fold increase in $v_s$ leads to a ca. 8% increase in $F_s$), is a clear signature of boundary lubrication, i.e. sliding dissipation occurs via molecular contacts at the slip plane rather than any viscous effect[40].

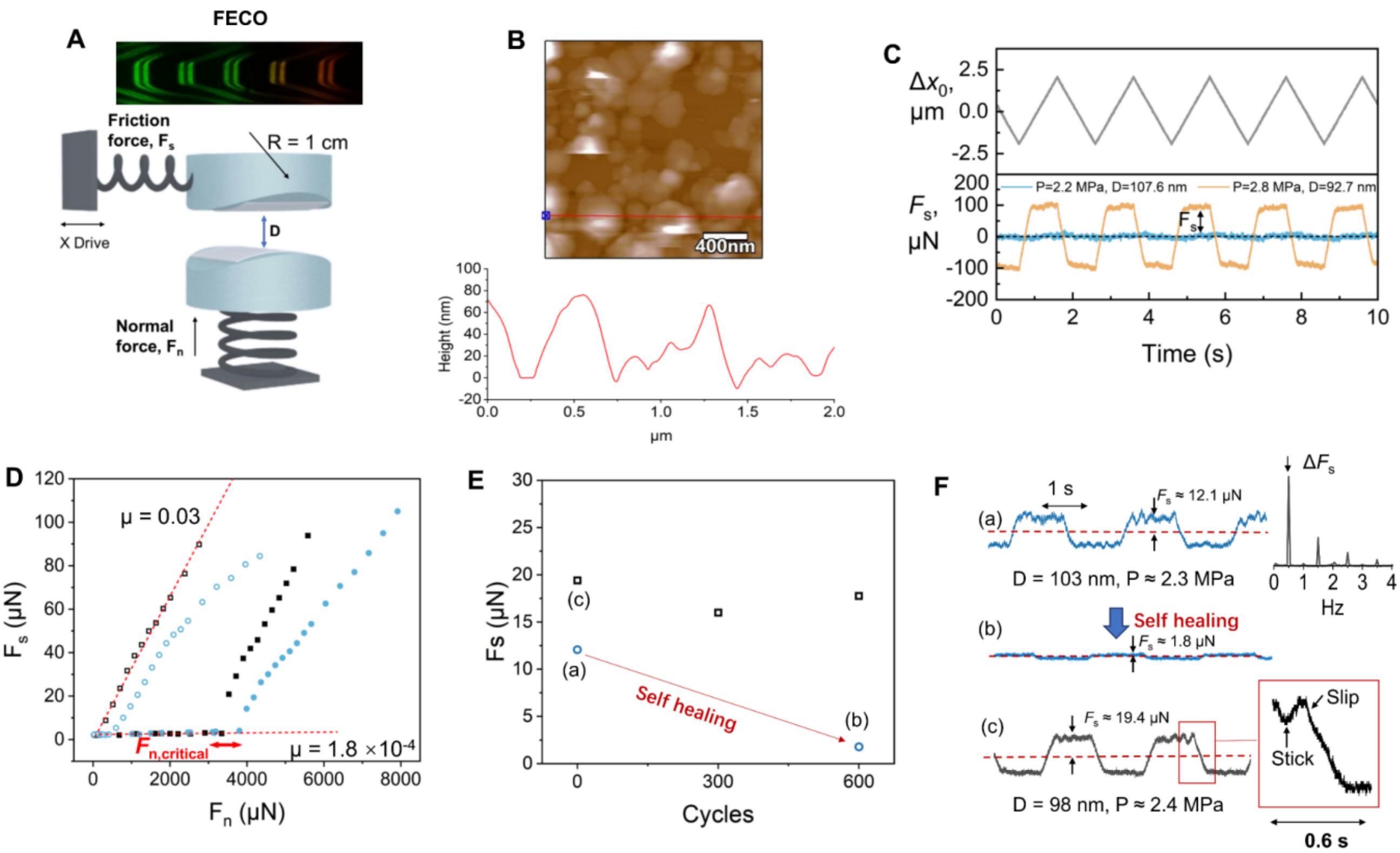

**Figure 3**. Interfacial sliding dissipation by mica-adsorbed NLG films. (A) Schematic representation of the SFB for normal and friction force measurements between curved surfaces[41] (mean radius of curvature $R$), showing (top) an example of fringes of equal chromatic order (FECO), used to determine the surface separation and inter-surface forces (see SI section 3). (B) AFM morphology of the NLG layer on mica with height cross sections below. (C) Representative applied lateral motion $\Delta x_0$ to the upper surface (top) and shear force $F_s(t)$ traces (bottom) across NLG dispersions. (D) Variation in $F_s$ *vs* $F_n$ for confined NLG layers (n=2), showing the critical load $F_{n,critical}$ for H-bond disruption determined from the normal force profiles in Figure 5C. (E) Shear force ($F_s$) as a function of sliding cycles for NLGs under different applied pressures, and (F) representative traces. The variation of $F_s$ with sliding velocity $v_s$ is shown in Figure S5. The corresponding pressure ($P$) values are from the Hertzian expression $P = F_n/(\pi*(F_n*R/K)^{2/3})$, where $F_n$ is the normal force, and the effective mica/glue modulus $K \approx 3 \times 10^9$ N/m$^2$ [8, 42].

Typical shear-force $F_s(t)$ vs. time ($t$) traces for the LUV and PTN layers are shown in Figure 4A and B. Unlike NLGs, LUVs ruptured on adsorption to the negatively-charged mica, and fused to form a lipid bilayer of thickness ~6 nm on the mica substrate (Figure 4C), in line with earlier observations[14]. We attribute this to the adsorption of the PEG moieties to the substrate, as well as bridging by the $Ca^{++}$ of the negatively-charged DSPE headgroups to the mica, leading to localized bilayer stress and rupture, as indicated schematically in Figure S6 and analyzed more quantitatively in the SI (SI section 7). This may be of particular relevance when NLGs are used for drug delivery in the synovial cavity, recalling that the articular cartilage surface is also negatively charged[15] and indeed with a surface potential similar to that of mica. PTN alone, notably, was not able to efficiently bind to the mica substrate and was readily removed by the AFM tip on scanning (Figure S7); this may be attributed to its residual negative charge after aggregation (a process which consumed the divalent ions in solution, see S8 in SI), and the resulting repulsion from the negatively charged mica. As shown in Figure 4D, the LUV layer reveals coefficients of friction (COF, $\mu = F_s/F_n$) in the range of 4.1-5.9 × $10^{-4}$ at pressures up to ca. 7.6 MPa on first approaches, with systematically lower μ (1.3-2.4 × $10^{-4}$) on second approaches at a given contact point. Such low friction coefficients of LUV boundary layers (as well as their reduction on second approach due to removal of loosely-attached vesicles) are very similar to earlier studies[14]. They are attributed to hydration lubrication at the slip-plane between the highly-hydrated head groups exposed on the surface of lipid bilayers, where water molecules tightly bound to headgroups retain a high shear fluidity through rapid relaxation, and thus achieve very low interfacial dissipation (and COF) on sliding.[6, 43] For the case of the PTN (Figure 4B) the shear force variation in the $F_s(t)$ traces, where neither of the lower $F_s(t)$ traces achieves a flat sliding plateau, may be attributed to the weak attachment of the PTN on the mica indicated in the $F_n(D)$ profiles (see details in the following section; fig. 5B), as well as the possibility of internal hydrogel bond rupture on shear. These would allow some sliding to occur; but as no plateau in the $F_s(t)$ traces is achieved (Figure 4B bottom), sliding friction force cannot be measured.

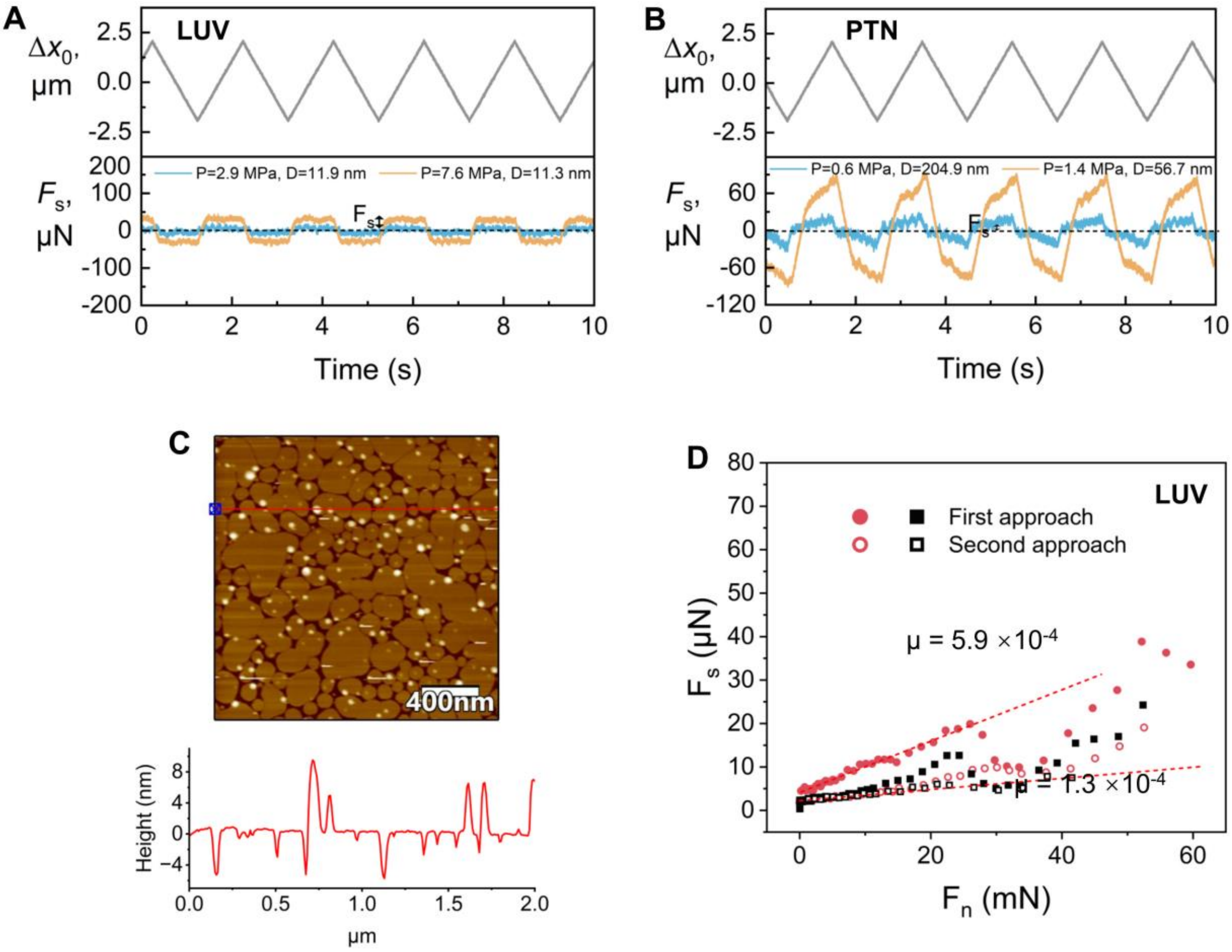

**Figure 4**. Interfacial sliding dissipation by mica-adsorbed LUV and PTN layers. Typical shear (or friction) force $F_s$ versus time traces taken directly from the SFB. The upper traces in each figure (A: LUV; B: PTN) show back-and-forth lateral motion $\Delta x_0$ applied to the upper mica surface, transmitting lateral (frictional) forces $F_s$ (lower traces) to the lower surfaces at different compressions. (C) AFM images of the mica surfaces bearing LUV with height cross sections below. (D) Variation of $F_s$ *vs* $F_n$ for confined LUV layers.

**Surface Interactions**

To better understand the load-responsive lubrication mechanism, the normalized force profiles in the Derjaguin approximation, $F_n(D)/R$ *vs.* surface-separation $D$ between approaching mica surfaces bearing LUVs, PTNs and NLGs are measured. As shown in Figure 5A, a short-ranged repulsion onsets at a separation of approximately 70 nm in the salt dispersion of LUVs, arising from repulsive steric interactions likely due to loosely-attached intact LUVs on the bilayers of the ruptured bilayers on each surface, as earlier observed[14]. At the strongest compressions, following removal of any loosely-attached LUVs by shear, the interactions reached a “hard-wall” repulsion at $D_{HW}$ = 10 ± 2 nm, corresponding to one lipid bilayer on each surface, characteristic of the ruptured vesicles seen in fig. 4C. In contrast, in PTN dispersions (Figure 5B) a weak, long-ranged repulsive force commences from ca. 300 nm, increases slowly, and is sometimes followed by a jump-to-contact to a final distance of 2.3 ± 0.8 nm at certain contact points, indicating essentially complete removal of the PTN from between the surfaces. At other contact points, however, no jump is observed, attributed to trapping between the surfaces of the weakly adsorbed PTNs.

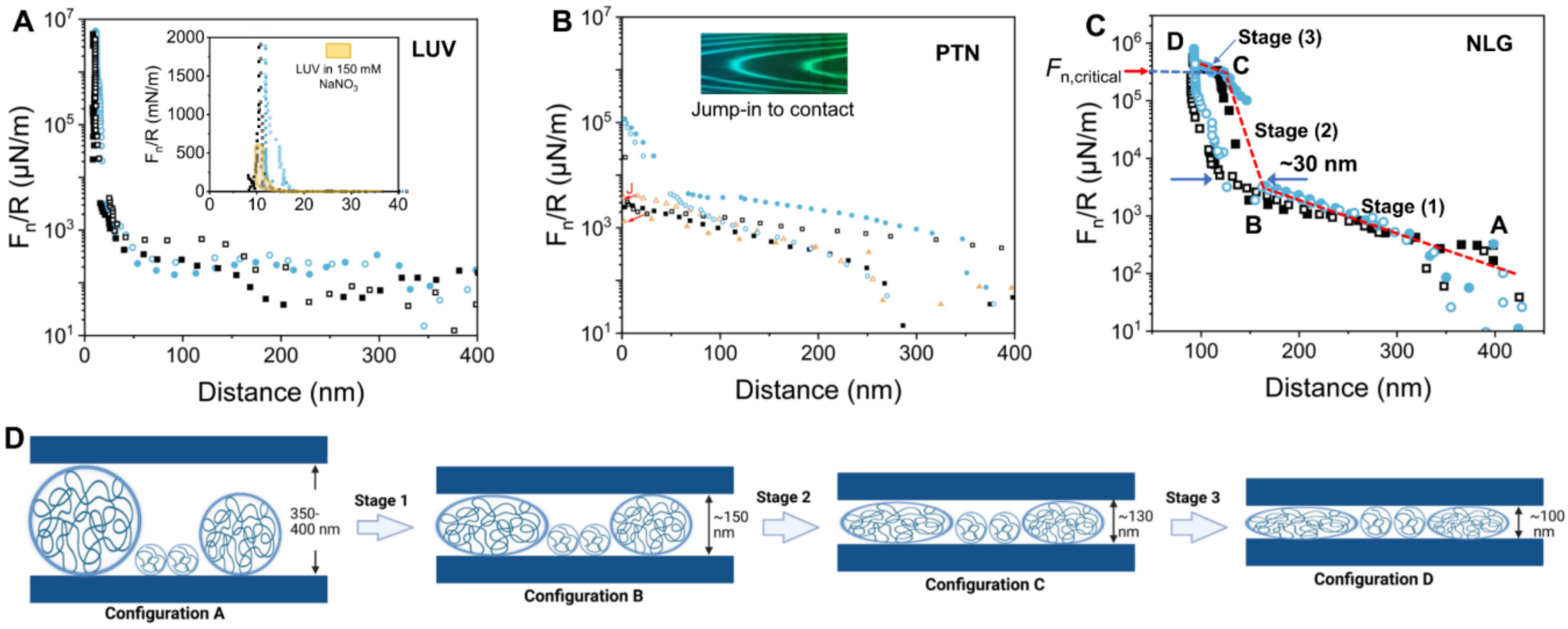

**Figure 5**. Normalized SFB-measured force profiles $F_n(D)/R$, based on independent separate experiments, between (A) LUV-, (B) PTN- and (C) NLG-coated mica surfaces across dispersions of the respective species in 2 mM $Ca(NO_3)_2$ solutions. In (C) the configurations A – D and the stages 1 – 3 are discussed in the text and illustrated in panel D. $D = 0$ nm is defined as bare mica-mica contact in air. Filled and empty symbols are for first and for subsequent approaches. (D) Illustration of the confined NLGs layer in different confinement stages (based on fig. 5C). See text for details.

The behavior of NLG, which is of greatest interest in the context of this study, on first approaches (Figure 5C) may be roughly characterized in 3 progressive stages between 4 configurations as follows, and as schematically illustrated in Figure 5D and considered in detail below.

**(1)** configuration A in Figure 5D corresponds to $D \approx 350 - 400$ nm where the opposing surfaces start to feel steric repulsion due to the largest NLGs (Figure 1E), and in stage 1 a long-ranged monotonic repulsion between the NLG layers on the approaching surfaces is seen, and the separation reduces to $D \approx 150$ - 160 nm (configuration B). This repulsion increases roughly exponentially with decreasing $D$, Figure 5C, reminiscent of counterion osmotic pressure in the classic Poisson-Boltzmann model[44]. However, the observed apparent decay length (i.e. the Debye screening length, see SI, Section 9) of ca. 80 nm is very much larger than the value of ca. 3.9 nm expected at the 2 mM Calcium ion concentration of the aqueous medium. Thus we conclude that this long-ranged repulsion in stage 1, 350 – 400 nm down to $D \approx 150$ - 160 nm, is due not to counterions but rather to steric repulsion arising from progressive distortion and flattening to a uniform thickness of the NLG particles, whose unperturbed size is quite polydispersed and may reach some hundreds of nm (Figure 1E). **(2)** At $D \lesssim$ 150 - 160 nm, Figure 5C configuration B, a sharper rise in $F_n$ is seen as the surfaces approach, stage 2, down to $D \approx 125$ nm. We attribute this to compression of the more uniform compressed NLG layer (once the largest NLG particles have been compressed to a size similar to the mean NLG dimensions), as water is further squeezed out of the nanogels, increasing the local osmotic pressure[45] and leading to configuration as in Figure 5D. **(3)** In the compression range 125 nm $\gtrsim D \gtrsim$ 95 nm, stage 3, the behaviour is seen to change again, and we attribute this, as detailed below, to rupture/rearrangement of hydrogen bonds between TA and PVA molecules within the nanogels, and the possible disruption of the surrounding membrane lipids from the nanogel core due to rupture/rearrangement of the nanogel/lipid H-bonding. This leads to the final 'hard-wall' configuration as schematically shown in Figure 5D. **(4)** Finally, as the surfaces coated with NLGs are further compressed, there is a "hard wall" separation at $D_{HW} \approx 90$ nm (i.e., the limiting separation at the highest loads applied), as illustrated in Figure 5D configuration D; we ascribe this to near-solid-like repulsion by some of the largest NLGs from which the water has been almost entirely been squeezed out, so that they become nearly incompressible. This is consistent with the observation that initial steric contact between the largest adsorbed NLG particles (stage (1)) occurs at around 350 - 400 nm, and assuming that some 75% of the NLGs is water (as indicated by thermogravimetric measurements, Figure S2), compression to 25% of this initial separation (i.e. from 350 - 400 nm to ca. 90 – 100 nm, as observed) would indeed squeeze nearly all of the water out of these largest NLG particles.

On separation of the surfaces followed by a second approach at a given contact point, empty symbols in Figure 5C, we observe a long-ranged repulsion similar to stage (1) of the first approach, save that the range of stage (2) is some 30 nm closer in than on first approach, as indicated in Figure 5C, leading directly to the hard-wall configuration, while stage (3) – H-bond rearrangement - is no longer observed at all. That is, we observe an irreversible reduction in NLG thickness of some 30 nm due to H-bond rupture/rearrangement on compression at $F_n > F_{n,\,critical}$. We note that the critical load at which this irreversible change commences is in the range $F_{n,critical} = 3.5 \pm 1$ mN, corresponding to a mean contact pressure ca. 2.5 MPa. It is of special interest, and supportive of our interpretation of irreversible distortion, that this pressure is similar to the mean energy density of H-bonds in the enclosed nanogel (SI, section 10). At pressures up to this value one expects elastic (reversible) response of the nanogel to the stress, but at higher pressures, i.e. at loads $F_n > F_{n,critical}$, one may expect rupture of the H-bonds. This behaviour is fully consistent with our interpretation of stage (3) above, confirming that the deformation through H-bond rupture/rearrangement of the dynamic nanogel network, which occurs in stage (3) on first approach, is irreversible, and leads to an average decrease of ca. 30 nm in the NLG dimension. We emphasize that even in the hard-wall region (configuration D in fig. 5D, where the surface separation is ca. 100 nm) a large fraction of the NLGs remain within their unperturbed dimensions, as shown in the DLS profiles (fig. 1E) where the peak in the NLG size distribution is also around 100 nm. This implied that any drug loading of these (~ 100 nm) vesicles would still be fully available as these NLGs would not have been compressed.

**Coarse-grained dynamics simulations**

Coarse-grained molecular dynamics (CGMD) simulations were performed to examine the above hypothesis that high pressure induces irreversible changes in the internal H-bond network through H-bond rupture and rearrangement. The structures of TA, PVA and water molecule were simplified into CG models, composed of 43, 10 and 1 microspheres, respectively (Figure 6A).[46] The model of the interior PTN core, consisting of 4000 water, 100 PVA and 5 TA CG molecules, was constructed in a simulation box of 80 Å × 80 Å × 80 Å. After equilibration in the isothermal–isobaric (NPT) ensemble with a time step of 1 fs at 300 K, uniaxial compression was applied to calculate the compressive stress–strain response (Figure 6B). The snapshots (Figure 6C) are the corresponding stress–strain states showing the network evolution using OVITO software[47]. As shown in Figure 6B, the compressive stress–strain curve is approximately linear up to a strain of ~9%, corresponding to the elastic regime. Beyond the linear (elastic) region, some deformation becomes permanent and irreversible, i.e. plastic deformation. Compressive stress reaches a maximum at $\varepsilon \approx 0.24$, after which it gradually decreases, suggesting softening of the H-bonded network, consistent with stage 3 in the *normal* force profile (Figure 5C). In this regime, rearrangement of hydrogen bonds is proposed to facilitate lubrication recovery by promoting redistribution of water and lipids within the slip plane and re-formation of a highly lubricated interface between the hydrated headgroups of intact, continuous lipid bilayers, consistent with the decrease in friction force under sliding shown in Figure 3E (blue symbols). However, the deformation due to the rupture/rearrangement of H-bond networks is reversible, resulting in the closer second approach. Upon further compression, the network stiffens, the *normal* force profile reaches the "hard wall" and lubrication recovery becomes less pronounced (dark symbols in Figure 3E), likely because reorientation of the surrounding lipid bilayer is increasingly restricted. While it is important to be aware of the limitations of such simulations, both because of the coarse-graining and because of the limited sample size, it is nonetheless of interest that plastic deformation in the MD simulations starts at close to 2 MPa stress, in line with both the SFB results above, fig. 5C, and the estimated H-bond energy density within the nanogel (SI, section 10). Likewise, the extent of strain ($\varepsilon \approx 0.24$) to the maximal stress in the MD simulations is comparable to that indicated by the SFB force profiles, see fig. 5C where the corresponding ~30 nm compression between points B and C represents a strain $\varepsilon \approx 0.2$, though the NLG size dispersity means this is an average value over a range of particle sizes. These approximate agreements between experiment and the MD simulations, provide grounds for believing that, despite their limitations, the latter do capture the main features of the NLG behaviour on compression.

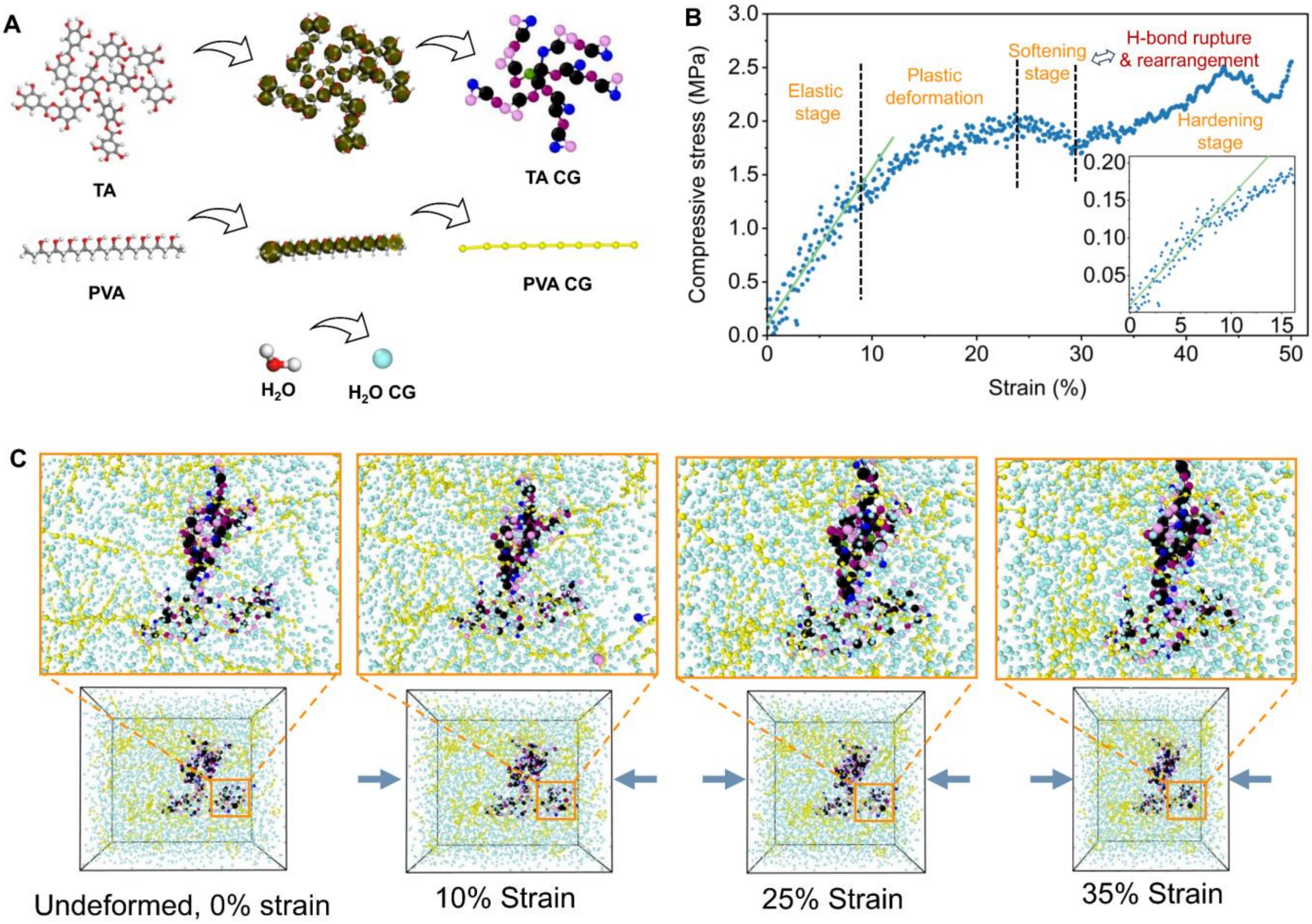

**Figure 6**. CGMD simulations of H-bond network containing 75wt% water. (A) CG models of TA, PVA and water molecules. (B) Representative compressive stress-strain curve from MD simulations,

showing the behaviour of PTN network. (C) Snapshots of PTN network at 0 % (undeformed state), 10 %, 25 % and 35 % compression strain are applied, showing the compressive direction and movement of CG molecules.

**Robustness of bilayer/nanogel attachment**

Under load and sliding, lipids may detach from the nanogel, potentially disrupting their ability to reconstruct the low-friction, exposed lipid bilayer surface. To examine directly whether or not such detachment occurs, and to what extent, we carried out the following measurements. The interior nanogel and lipid bilayer of the NLG were labeled with rhodamine B (RhB) and DiD, respectively, and the labeled NLGs, dispersed in 2 mM $Ca(NO_3)_2$ solution, were adsorbed onto a mica surface (Figure 7A). The wear behavior of the NLGs was examined against a stainless steel counter-surface across 2 mM $Ca(NO_3)_2$ solution under a load of 1 N using a UMT tribometer; from the width of the wear track, this corresponds to a mean contact pressure of $P \approx 15$ MPa, which is comparable to the highest measured contact pressures in human articular cartilage[37]. The confocal analyses were used to assess spatial association between the interior nanogel and lipid bilayer of the NLGs (Figure 7B), in order to assess to what extent they were detached from each other on shear. Rectangle region of interest (ROI)-based analysis of the confocal images showed a positive spatial association between the DiD-labeled lipid bilayer and RhB-labeled interior nanogel, with a high Pearson's correlation coefficient (PCC) = 0.80 ± 0.09 prior to shear, and a PCC value = 0.72 ± 0.05 (n=4; see details in SI, Section 11) in the wear track following shear. In addition to this rectangle ROI-based analysis, we also carried out a line-profile analysis across the non-wear and wear regions, which revealed coincident fluorescence peaks between the DiD-labeled lipid bilayer and the RhB-labeled nanogel (Figure 7C, D). The non-wear region again exhibited a high PCC value (0.81 ± 0.05; Figure 7E), again indicating consistently strong DiD/RhB line-profile correlation and spatial association between the lipid- and nanogel-associated signals. In the wear region, although sliding increased the heterogeneity of lipid–nanogel co-distribution, as indicated by a decreased PCC value (0.68 ± 0.25), the two signals remained quite strongly positively associated. This clearly indicates that, while there is some decrease in association between the nanogel and lipid bilayer following shear under normal stress as revealed by the slight decrease in PCC, most of the lipid bilayers are retained on the enclosed nanogel. This retention of lipids on nanogels is attributed to the strong interactions (i.e., H-bonds and cation-π interactions) between lipid headgroups and exposed OH groups/aromatic rings on nanogel surfaces, as also evidenced (fig. 1B, C) by the FTIR and XPS results. This also ties in with our observation (fig. 3E) that when compressed and sheared at loads slightly beyond the threshold ($F_{n,critical}$, $P \approx 2.3$ MPa), in the region C => D in fig. 5C, ruptured lipid bilayers can recover (fig. 3E) to form continuous hydrated layers under shear (Figure 7F), due to their fluidity as indicated in fig. 2A. Upon further compression, however, at loads higher than point D in fig. 5C, the nanogels undergo irreversible H-bond rupture and become stiffer; this hinders lipid movement, resulting in diminished lubrication recovery; these stages are schematically illustrated in fig. 7F.

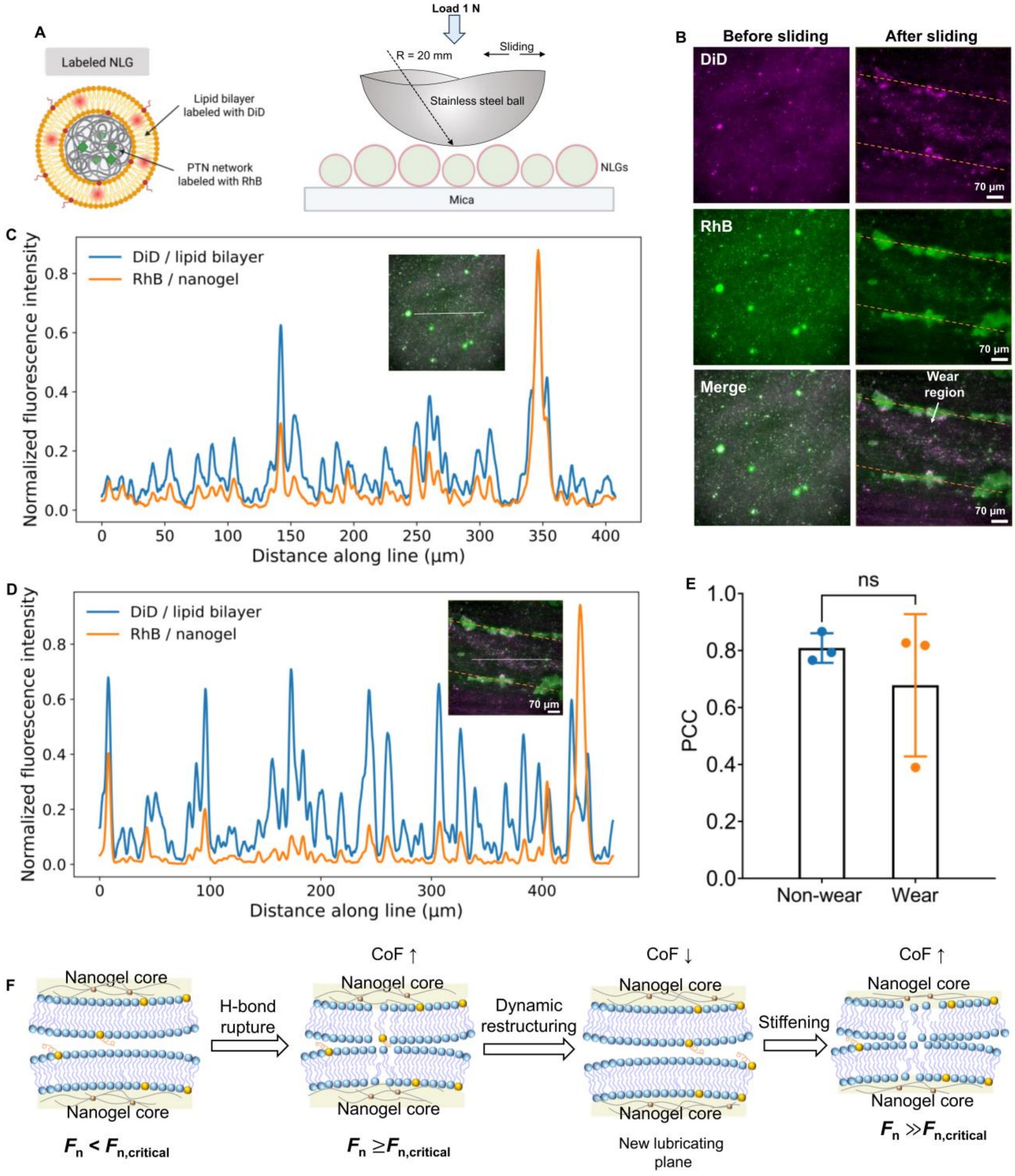

**Figure 7**. Detachment of lipids from nanogels as imaged by confocal microscopy. (A) The schematic diagram of RhB/DiD labeled NLG and sliding configuration. The lipid bilayer was dyed with DiD, and the interior nanogel was dyed with RhB. (B) Confocal images of NLGs adsorbed on mica before and after sliding under a load of 1 N (mean pressure $P \approx 15$ MPa). Larger particles may represent contaminants or aggregated NLGs from centrifugation. Line-profile analysis was performed across the non-wear (C) and wear (D) regions (white line in the insert image), and the fluorescence intensities of both channels were normalized to compare the spatial coincidence of RhB and DiD intensity peaks. (E) Line-profile PCC between DiD- and RhB-labeled NLG components in non-wear and wear regions. Individual points represent separate line profiles, and square markers with error bars indicate the mean ±SD. (F) Schematic illustrating the proposed origin of the variation in frictional dissipation as the load increases from below to high above its critical value $F_{\mathrm{n,critical}}$ (see also text).

**Potential for Drug Delivery**

To assess the potential of the system for clinical application, we evaluated its *in vitro* cytotoxicity against HEK293 and HeLa cell lines using an MTT colorimetric assay. HeLa and HEK293 cells were cultured in 96-well plates and typically reached confluence within 24 h. The cells were then co-incubated with either LUV or NLG solutions for 24 h, after which they were subjected to an MTT assay (Figure S9). The MTT assay demonstrated that both LUV and NLG were non-toxic to HEK293 and HeLa cells, with cell viability exceeding 80% even at concentrations up to 1000 µg/mL, indicating good biocompatibility (Figure 8A, B). The drug loading capacity (DLC%) and entrapment efficiency (EE%) of NLGs were measured using RhB as drug model. The amount of non-entrapped RhB remaining in the combined supernatant and washing solutions was quantified by UV-Vis spectroscopy at 560 nm, allowing the amount of encapsulated RhB to be determined indirectly (see details in SI, Section 13). The RhB-loaded NLGs showed a DLC of 6.9 ± 0.2% and an EE of 96.9 ± 0.5% (n = 3), well comparable with levels in other drug-loaded nanoparticle systems[48], indicating that the co-precipitation process enabled effective incorporation of RhB into the nanogel system.

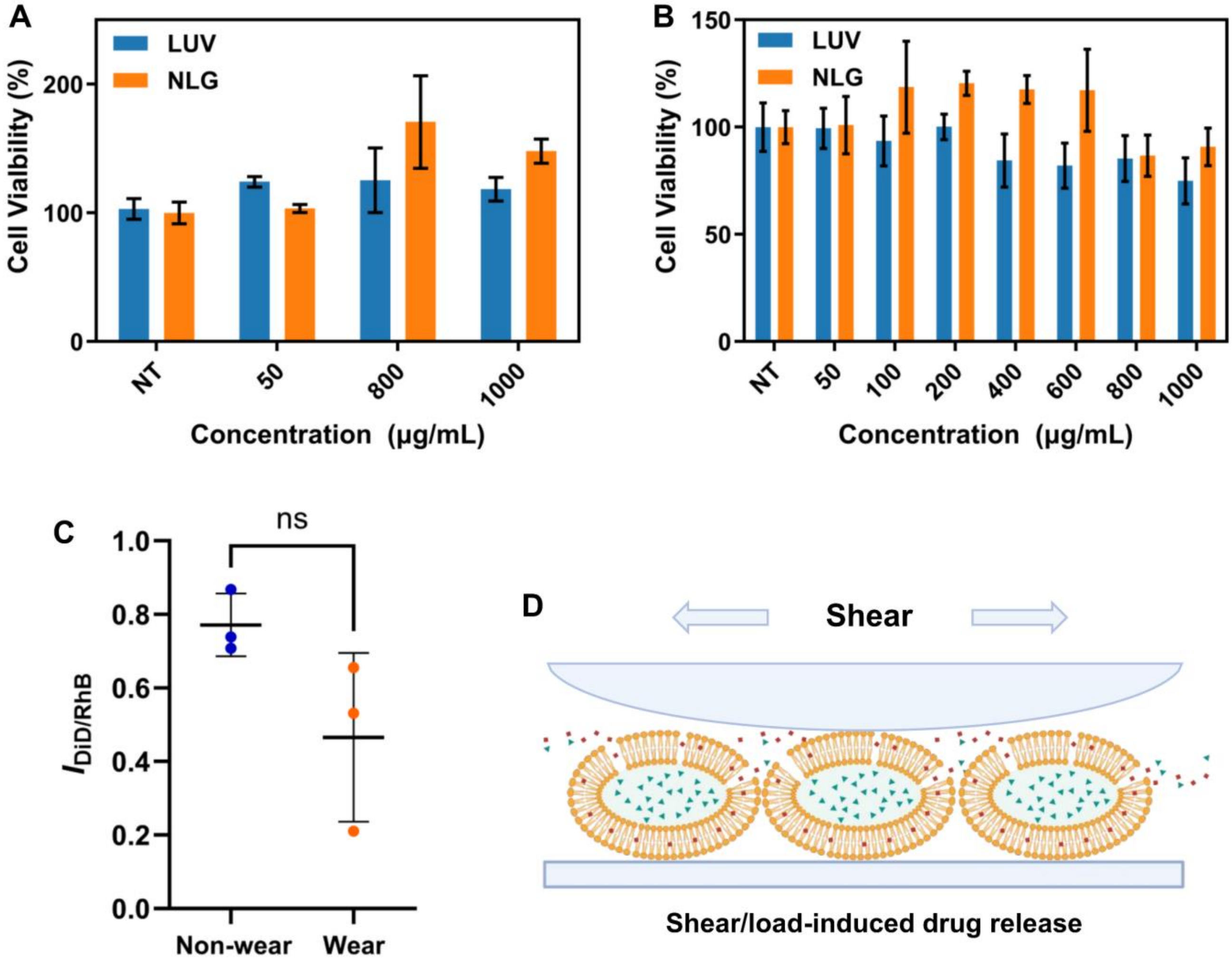

**Figure 8**. Cytotoxicity analysis was performed using the MTT assay. HeLa and HEK293 cells were incubated with LUVs or NLGs for 24 h at 37 °C. Quantification of cytotoxicity in HeLa (A) and HEK293 (B) cells. Percentages are corrected based on the viability control (culture medium alone). Abbreviation: NT, untreated cells. (C) DiD/RhB intensity ratio ($I_{DiD/RhB}$) in non-wear and wear regions (n=3, see details in SI, section 11). (D) Illustration of shear/load-induced drug release.

The drug retention in NLGs after sliding was tested by a tribometer, using DiD as a hydrophobic drug model in the lipid bilayer and RhB as a hydrophilic drug model in the interior nanogel (Figure 7A). Figure 7B shows that both DiD and RhB signals remained after sliding under ~15 MPa, indicating good retention of drugs in both lipid bilayer and interior nanogel. The intensity ratio of DiD/RhB ($I_{DiD/RhB}$) in NLGs before and after sliding are calculated from raw, background-corrected fluorescence intensities (Figure 8C; see details in SI, Section 11). The $I_{DiD/RhB}$ decreased from 0.83 ± 0.08 in the non-wear region to 0.52 ± 0.20 in the wear region, indicating a reduction in lipid-associated fluorescence relative to the RhB-labeled nanogel signal after loading and sliding. We attribute this to the fact that sliding-induced rupture/reorganization of the lipid bilayer promoted DiD release, while the interior nanogels, many of which were not perturbed by the load (see fig. 5D and associated earlier discussion), provided stronger retention of RhB. The shear/load induced drug release is illustrated in Figure 8D.

## Conclusions

To summarize, we find that NLGs, comprising nanogel networks crosslinked by H-bonds, and encapsulated by sterically-functionalized PC lipid bilayers to which they are also H-bonded, provide very efficient lubrication, followed by an unusual bond rupture/reconstruction behaviour when progressively compressed and sheared between sliding surfaces. In contrast to PEG-stabilized PC liposomes alone, which spontaneously rupture to form bilayers on negatively charged surfaces, the NLGs form topologically-stable surface assemblies on such surfaces (we recall in passing that most biological surfaces, including articular cartilage, are likewise similarly-negatively-charged[15]). Up to moderate compressions (ca. 2-2.5 MPa, comparable with mean physiological pressures in human synovial joints[37]) the NLGs remain intact as the nanogel network deforms elastically, and exhibit very weak interfacial sliding dissipation, with friction coefficients (reflecting the dissipation, eq(1)) as low as $\mu \approx 10^{-4}$. This low value is attributed to hydration lubrication at the slip-plane of the highly-hydrated continuous lipid headgroups of the encapsulating PC bilayer. Beyond this critical pressure, which is comparable to the H-bond energy density within the nanogels (SI, section 10), the dissipation approaches to a (still-low) value $\mu \approx 10^{-2}$. This scenario is considerably strengthened by our observation that the load at which the friction coefficient increases on a first approach is very close to the critical load $F_{n,critical} \approx 3.5\pm0.5$ mN at which irreversible H-bond-rupture is attributed to occur in our configuration, based on the *normal* force profiles (fig. 5C). Consistent with this, as noted earlier, is that this load corresponds to a normal contact stress (2 – 2.5 MPa) similar to the H-bond energy density (SI, section 10), so that higher stresses lead to H-bond rupture. Further support for this scenario comes from the stress decrease observed during compression in the compressive stress–strain curve obtained from the CGMD simulations (fig. 6B). We attribute the increase in CoF to H-bond rupture within the largest (and hence most compressed) nanogels, and between them and the surrounding lipid bilayer, in response to pressures higher than the H-bond energy density within the nanogels. The resulting damage to the originally-continuous lipid bilayer at the slip plane results in more interfacial dissipation with the opposing boundary layers, and hence to increased CoF on sliding. However, at pressures slightly above this threshold (stage 3 of the *normal* force profile, fig. 5C), dissipation gradually decreases and returns to $\mu \approx 10^{-4}$ after ~600 sliding cycles. In this regime, the softened network (as revealed by CGMD) facilitates the redistribution of water and lipids within the slip plane, resulting in the re-formation of a continuous hydrated lipid layer.

Some remarks concerning the applicability of NLGs for combined cartilage lubrication and drug delivery are in order: this first examination of NLGs as lubricants/drug-carriers is simplified, using a pure water medium rather than the complex synovial fluid surrounding cartilage, with its high ionic strength (~150 mM NaCl) and many macromolecules (e.g., albumin, γ-globulin, hyaluronic acid (HA)), which may affect interfacial dissipation. We recall however that earlier work has clearly shown that boundary lubrication by lipid head-group layers is little-affected by high salt concentration[49] or by HA[50]. Likewise, while SFB measurements cannot fully replicate the intricate contact mechanics and dynamic movements of natural joints, they do reveal the remarkable boundary-lubrication properties of the NLGs, arising from their exposed lipid bilayers; such boundary lubrication is believed to be the main mechanism underlying the remarkable lubricity of healthy articular cartilage[3]. Parameters such as nanogel modulus, cross-linking density, size dispersity and lipid bilayer composition (e.g., unsaturated phosphatidylcholines, fatty acids, or proteins) could further modulate interfacial structure and lubrication. Future work should incorporate more biomimetic systems to build on these insights. Overall, this study elucidates lubrication mechanisms, drug-loading and biocompatibilty by nanoparticles comprising lipid bilayers supported by supramolecular networks, laying groundwork for simultaneous friction-reducing therapies and targeted drug delivery.

## Experimental Section

### Materials

1,2-distearoyl-sn-glycero-3-phosphoethanolamine-N-[methoxy(polyethylene glycol)-5000] (ammonium salt) (18:0, DSPE-PEG, PEG $M_w$ of 5000 Da) was purchased from Avanti Polar Lipids. Hydrogenated soy phosphatidylcholine (HSPC) was purchased from Lipoid (Ludwigshafen, Germany). Tannic acid (TA, ACS reagent), poly(vinyl alcohol) (PVA, $M_w$ 89,000~98,000, 99+% hydrolyzed), calcium nitrate hydrate (99.997% trace metals basis), rhodamine B (RhD, ≥95%, HPLC) and MTT (thiazolyl blue tetrazolium bromide) were purchased from Sigma-Aldrich. AF568 carboxylic acid and DiD were purchased from Lumiprobe. Water was highly purified by a Barnstead Nanopure system and had resistivity of 18.2 MΩ and total organic content of ca. 1 ppb.

**Preparation of LUVs**

PEGylated, large unilamellar vesicles (LUVs) were made via a film hydration-extrusion technique[14]. Briefly, HSPC (46.7 mg)/DSPE-PEG (5.3 mg, 1.5 mol%) was dissolved in methanol and chloroform (1 mL, 1:1 v/v). After that, organic solvent was removed by nitrogen overnight. The lipid film was then hydrated with pure water at 70 °C to reach a concentration of 6 mM. The resulting multilamellar vesicles were downsized to form LUV by stepwise extrusion (Northern Lipids, Burnaby, Canada) through polycarbonate membranes starting with 400 nm (11 cycles), then 200 nm (11 cycles).

**Preparation of PTNs**

PVA/TA nanoparticles (PTNs) were obtained by coprecipitation.[51] Briefly, an aqueous solution of PVA (50 mg, 5 mL) was added to an aqueous solution of TA (15 mg, 5 mL). The mixture, which turned white immediately (fig. S1), was sonicated at room temperature for 15 minutes in an ultrasonic bath. The excess of reactants was eliminated by dialysis against water for 48 hours to obtain a white suspension of nanoparticles.

**Preparation of NLGs**

PTN@lipid bilayer nanolipogels (NLGs) were prepared using a thin-film hydration method, as previously reported.[21] The lipid film, whose preparation is similar to LUVs, was sonicated with 2.5 mg of PTN in 10 mL of water. The resulting NLG were passed through a 400 nm polycarbonate membrane (11 cycles) using an extruder.

**Drug loading**

PTNs loaded with RhB were prepared by co-precipitation. Briefly, 50 mg of PVA was dissolved in 5 mL of deionized water containing 1.3 mg of RhB, and this solution was mixed with 5 mL of an aqueous TA solution (3 mg/mL). The resulting mixture was sonicated for 15 min, after which the nanogel pellet and supernatant were separated by ultracentrifugation at 150,000 ×g for 10 min. The pellet was redispersed in deionized water and subsequently coated with a lipid bilayer following the protocol described in the preparation of NLGs, yielding RhB-loaded NLGs. The RhB-loaded NLGs were then washed three times with fresh deionized water by centrifugation to remove any RhB loosely adsorbed on the surface. The supernatant collected after the initial ultracentrifugation step, together with all washing solutions, was combined and used to determine the amount of non-entrapped RhB by UV–Vis spectrophotometry at 560 nm using a standard calibration curve based on the Beer–Lambert law[52] (Figure S10; see details in SI, section 13). The purified NLGs were finally lyophilized to determine the weight of NLGs. The drug loading parameters were calculated as follows:

Drug loading capacity (DLC %) = Weight of RhD in NLGs/weight of NLGs × 100%

Entrapment efficiency (EE %) = Weight of RhD in NLGs/weight of initial RhD loaded × 100%

**Wear behavior and drug release of NLGs**

RhB-loaded PTNs were prepared as follows: 4.5. 50 mg of PVA was dissolved in 5 mL of deionized water containing 1.3 mg of RhB, and this solution was mixed with 5 mL of an aqueous TA solution (3 mg/mL). The resulting mixture was sonicated for 15 min, and then dialyzed for 3 days. DiD-labeled lipid films were prepared as follows: HSPC (46.7 mg), DSPE-PEG (5.3 mg, 1.5 mol%) and DiD (500 μg) were dissolved in methanol and chloroform (1 mL, 1:1 v/v). After that, organic solvent was removed by nitrogen overnight. The DiD-labeled lipid film was then hydrated with RhB-loaded PTN dispersion (2.5 mg/mL, 10 mL) at 70 °C and then extruded through a 400 nm polycarbonate membrane for 11 cycles using an extruder. RhB/DiD-labeled NLGs were finally obtained after washing with pure water by centrifugation. Mica sheets were incubated with the labeled NLGs in 2 mM $Ca(NO_3)_2$ solution for 2 h, then washed with 2 mM $Ca(NO_3)_2$ solution. The wear behavior of the labeled NLGs absorbed on mica was evaluated in 2 mM $Ca(NO_3)_2$ medium against a stainless steel ball (20 mm-diameter) using a CETR UMT tribometer (Bruker, MA, USA), with a reciprocating sliding mode (3 mm/s) at a load of 1 N for sliding 3 min, followed by confocal microscopy analysis (Leica, Germany). The DiD/RhB intensity ratios before and after wear were analyzed (see SI, section 11) to assess dye retention in NLGs under loading and sliding.

**Cell Culture**

HeLa cells were cultured in DMEM medium (Gibco) supplemented with 10% fetal bovine serum (FBS, Gibco) and 1% Penicillin/Streptomicin (Gibco) at 37 °C in a humidified incubator with 5% $CO_2$. HEK293 cells were cultured under the same conditions.

**Evaluation of Cytotoxicity**

Cytotoxicity was determined by the production of the purple formazan product upon cleavage of MTT by mitochondrial dehydrogenases in viable HeLa or HEK293 cells. In vitro dose-dependent experiments were carried out. The cells were seeded onto 96-well plates ($5 \times 10^4$ cells/well) in DMEM media. When the confluent state was reached (usually after 24 h), 50-1000 μg/mL of LUV or NLG in PBS solutions were then added. After 24 h incubation, the cells were incubated with 200 μL of MTT solution (0.5 mg/mL) for 3-4 h. The formed crystals were subsequently dissolved in 200 μL of DMSO and cell viability was quantified by measuring the absorbance at 570 nm using a multiwell-plate reader (Cary 100 Bio, Varian Inc., USA). The relative absorbance of non-treated cells was regarded as 100% viability. Each experiment was conducted two times, and three replicates were prepared for each concentration tested in every experiment. The potential biocompatibility of the different formulations tested was expressed as a viability percentage calculated using the following formula:

$$\%Viability = (ODtest/ODc) \times 100.$$

where ODtest was the optical density of those wells treated with the LUV and NLG solutions, and ODc was the optical density of the untreated wells. Cell cultures were carefully monitored before and during the experiments to ensure optimal cell density.

**Molecular dynamics simulation**

The coarse-grained molecular models were constructed using Moltemplate and Packmol on the Hongzhiwei Cloud platform (https://cloud.hzwtech.com), see also e.g. ref.[53]. Coarse-grained Moltemplate (.lt) files were prepared separately for TA, PVA and water. An amorphous polymer solution system was then generated with Packmol according to the prescribed composition, yielding a simulation box of $8 \times 8 \times 8$ $nm^3$ containing 5 TA molecules, 100 PVA molecules, and 4000 water beads/molecules. The interactions between coarse-grained beads were described using the MARTINI force field. All simulations were performed using LAMMPS[54]. Before dynamic simulation, the initial structure was first subjected to energy minimization using the conjugate gradient algorithm. The minimized system was then equilibrated under the NPT ensemble for 500 ps at 300 K and 0.1 MPa using the Nosé–Hoover thermostat/barostat. A time step of 1 fs was used during equilibration. The pressure in all three spatial directions was maintained at 0.1 MPa, and periodic boundary conditions were applied in all three dimensions. After equilibration, uniaxial compression was carried out along the z-axis. During the compression process, the time step was set to 0.0001 ps, and the system was compressed at a rate of 0.001 Å $ps^{-1}$. The corresponding stress–strain curves were recorded for mechanical analysis. The simulation trajectories and structural evolution were visualized and analyzed using the open-source software OVITO[47].

**Characterizations**

*(1) Attenuated Total Reflectance-Fourier Transform Infrared Spectroscopy (ATR-FTIR)*

ATR-FTIR analysis of the materials was performed using attenuated total reflection (ATR, diamond, Smart iTX) FTIR spectrometer Nicolet iS50 (Thermo-Fischer Scientific, USA, alignment

beam – He-Ne laser) having deuterated triglycine sulfate (DTGS/KBr) detector and KBr beamsplitter. Before analysis, the FTIR setup was purged with in-house nitrogen supply for ca. 3 h (flow rate is 20 sccm). Then the diamond crystal was cleaned with Milli-Q water and absolute ethanol and then wiped with optical leaning tissues. The background spectrum of the diamond crystal was taken before loading each sample. The atmospheric compensation spectrum (air) was taken before the analysis of each material and was further subtracted from the FTIR spectra of the materials where it was needed. A small piece of materials used without additional pretreatments was directly loaded into the ATR crystal and clamped. Each spectrum (autogain, resolution of 0.482 $cm^{-1}$ per point, range 525-4000 $cm^{-1}$) was collected 32 times and averaged. Each spectrum was taken twice. OMNIC spectroscopy software (Thermo-Fischer Scientific, USA) was used to resolve the spectra.

*(2) Dynamic Light Scattering (DLS)*

Size distribution and zeta potential of particles were characterized using a Zetasizer Nano-ZS instrument (Malvern Instruments Ltd., Malvern, WR, U.K.) with a scattering angle of 173° at 25.0 ± 0.1°C.

*(3) Differential Scanning Calorimetry (DSC)*

Typically, a 40 μl aliquot of lipid dispersion (30 mg/mL) was put in an aluminum pan. The DSC thermograms were recorded using a differential scanning calorimeter (DSC Q200, TA Instruments, USA) with a temperature ramp rate of 5 °C/min.

*(4) Fluorescence microscopy*

Fluorescence imaging was performed using Vutara SR352 microscope (Bruker, USA). Images were recorded using 1.3 NA 60× silicon oil immersion objective (Olympus) and Hamamatsu Orca Flash 4v2 camera with a frame rate at 50 Hz. Fluorescence images of nanolipogels were recorded using 560 nm and 640 nm excitation laser (maximal excitation of 6 $kW/cm^2$). Data were analyzed with the Image J software.

*(5) Cryogenic transmission electron microscopy (cryo-TEM)*

3.5 µL of the sample solution (1 mM) was applied to the carbon side of Lacey carbon grid (Cu, 300 mesh; Electron Microscopy Sciences, USA), which was glow discharged for 60 s at 0.5 mbar and 15 mA with Pelco easiGlow (Ted Pella, USA). Then the grids were blotted for 4 s from both sides at 4 °C and ~100% humidity and plunge-frozen into the liquid ethane cooled by liquid nitrogen using Vitrobot Mark IV (Thermo Fisher Scientific, USA). Cryo-EM images were collected using Talos

Arctica TEM, equipped with Falcon 4i direct detector (Thermo Fisher Scientific, USA) and operated at 200 kV. Image acquisition was performed using EPU 3.6 software (Thermo Fisher Scientific, USA) at a nominal magnification of 17.500× (pixel size 1.64 nm, underfocus 15-20 μm), 45.000× (pixel size 0.62 nm, underfocus 6-10 μm) or 120.000× (pixel size 0.118 nm, defocus 3-5μm) with a total dose not exceeding 100 e-/Å$^2$.

*(6) Atomic Force Microscopy (AFM)*

The morphologies and local Young's moduli of mica surfaces coated with particles under their dispersion (2 mM $Ca(NO_3)_2$) were measured using an MFP-3D stand-alone AFM (Oxford Instruments Asylum Research, Inc., Santa Barbara) at noncontact (tapping) mode. SNL tips (silicon tip on a silicon nitride cantilever, Bruker's SNL-10 probe with tip radius of 2 nm) were used with a nominal spring constant of 0.35 N/m. The AFM tip holder was irradiated in a UV-ozone cleaning system (ProCleaner Plus, BioForce Nanosciences, United States) for 15 min prior to use.

*(7) Thermogravimetric analysis (TGA)*

TGA was performed by a ThermoGravimetric Analyzer SDT Q 600. PTNs as prepared above (section 4.3) were deposited and collected by centrifugation at 10000 rpm, and excess surface water was removed using tissue paper. For each measurement, approximately 5~10 mg of PTN sample was placed in a clean TGA pan and heated from 40 °C to 500 °C at a heating rate of 10 °C/min under $N_2$ atmosphere. The mass loss below about 150 °C was interpreted mainly as the removal of physically adsorbed and bound water, while subsequent mass losses at higher temperatures were attributed to the thermal degradation of the PVA/TA polymer network.

**Author Contributions**

P.Z. and J.K. conceived the project and designed the experiments. P.Z. and A.M. carried out experiments. N.K. performed and characterized AFM imaging. A.S. performed and characterized ATR-FTIR measurements. R.K. performed and characterized cryo-TEM imaging. J.K. and P.Z. wrote the manuscript and analyzed the data. All authors participated in the discussion of results and revision of the manuscript.

**Conflict of interest**

The authors declare no competing interest.

**Acknowledgement**

We thank the European Research Council (Advanced grant CartiLube, grant number 743016), the Israel Science Foundation–National Natural Science Foundation of China joint research program (Grant 3618/21), the McCutchen Foundation and the Israel Science Foundation (grant 1229/20) for their support. We would like to thank Dr. Ulyana Shimanovich from Department of Molecular Chemistry and Materials Sciences for the supporting of ATR-FTIR instrument, Dr. Tali Dadosh from Department of Chemical Research Support for assistance on the STORM microscopy, Dr. Tatiana Smirnova and Dr. Yoseph Addadi from Department of life science core facility for assistance on the confocal microscopy, Dr. Yoav Barak from Department of Chemical Research Support (BioNANO lab) for microplate reader measurement assistance, Dr. Arina Dalaloyan from Department of Chemical and Biological Physics for assistance on in-vitro cytotoxicity evolution, Dr. Ayelet Vilan from Department of Chemical Research Support for assistance on XPS. Panpan Zhao wants to thank the Tom and Mary Beck Center for Renewable Energy as part of the Institute for Environmental Sustainability (IES) at the Weizmann Institute of Science. Graphical abstract was created using BioRender (https://BioRender.com).

**Supporting Information for:**

# Cytoskeleton-inspired, adaptive nanolipogels as superlubricating delivery vehicles

Panpan Zhao*[1], , Avijit Mondal[1], Nir Kampf[1], Aleksei Solomonov[1], Roman Kamyshinsky[2], Jacob Klein*[1]

1. Department of Molecular Chemistry and Materials Sciences, Weizmann Institute of Science, Rehovot, 76100, Israel
2. Department of Chemical Research Support, Weizmann Institute of Science, Rehovot, 76100, Israel

Emails: panpan.zhao@weizmann.ac.il, jacob.klein@weizmann.ac.il

**Table of Contents:**

## S1. Didital photos of dispersions

PTNs were well dispersed in pure water, but aggregated in 2 mM $Ca(NO_3)_2$ solution.

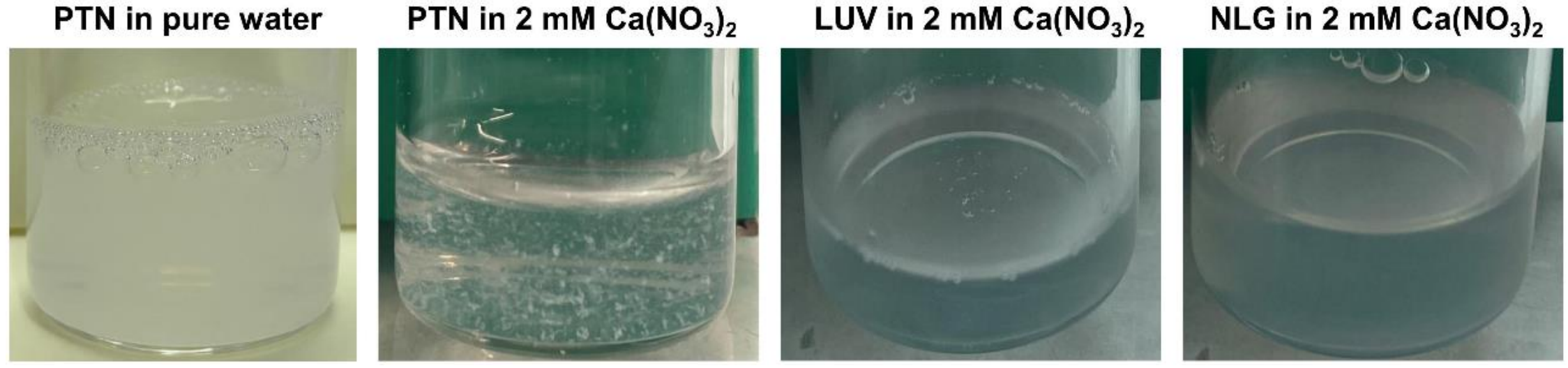

**Figure S1**. Digital photos of PTNs in pure water and 2 mM $Ca(NO_3)_2$, and LUVs, NLGs in 2 mM $Ca(NO_3)_2$.

## S2. Thermogravimetric (TGA) mearsurement

The plateau at 25% of original weight at just over 100 °C indicates that all water has evaporated, indicating an initial water content of ~75%. The subsequent weight losses at higher temperatures arise from thermal degradation of the PVA/TA network.

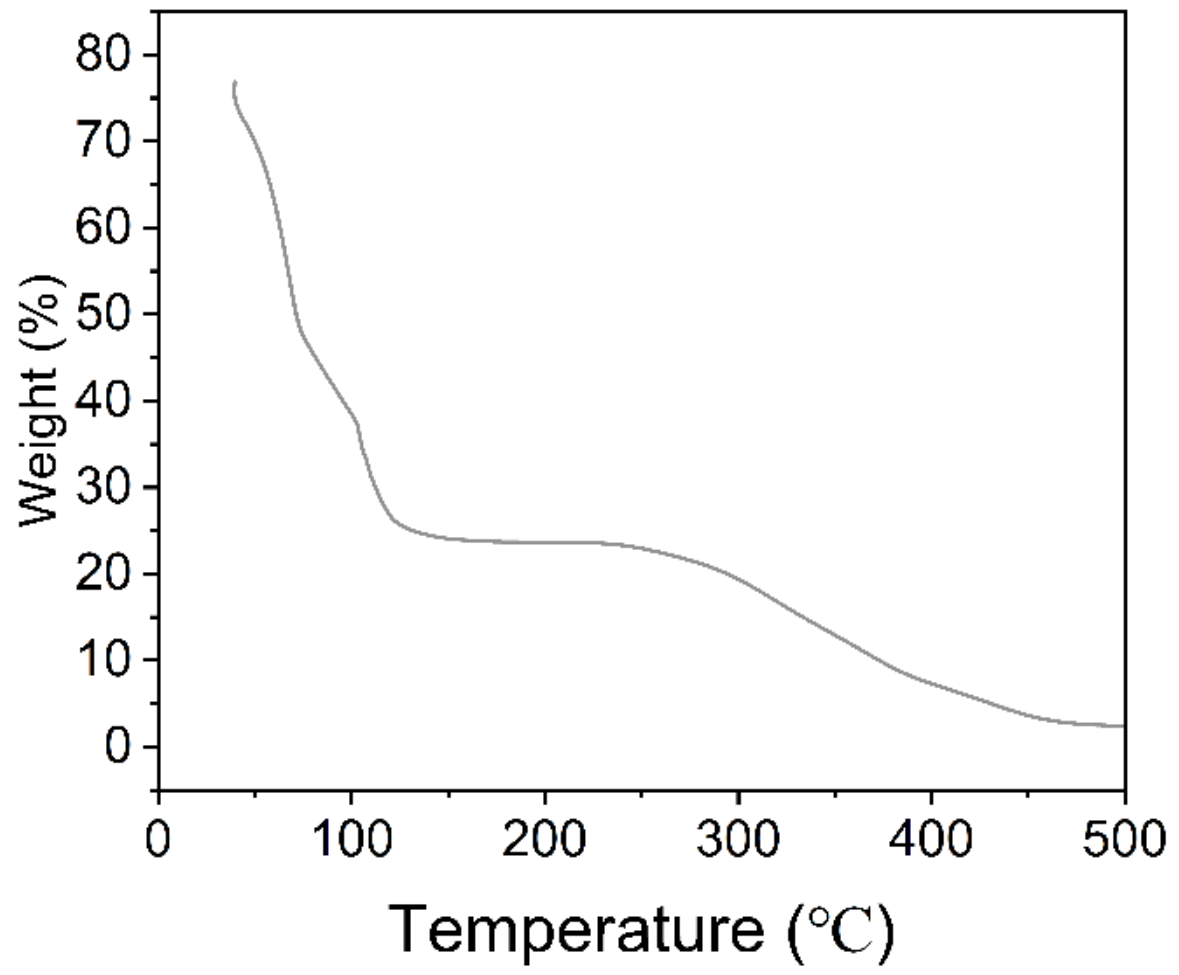

**Figure S2**. TGA curve of PTN in $N_2$ atmosphere.

## S3. Surface force balance (SFB)

The SFB setup and the principle for the measurements have been described in detail previously[1, 2]. Briefly, with reference to figure 2A: atomically-smooth mica sheets half-silvered on their back-side and coated with the desired species as described, are mounted on lenses in a crossed cylinder configuration – to eliminate alignment issues - as shown (figure 2A). This results in optical interference fringes of equal chromatic order (FECO) when white light is transmitted through them (upper inset to figure 2A shows spectrometric images of the FECO, from which the mean curvature $R$ of the mica sheets may be determined). The surfaces may be moved normally and laterally with respect to each other via a three-stage system, whose most delicate piezoelectric stage enables sub-nanometer motion control, while the FECO wavelengths yield the absolute separation D between the surfaces within a 2-3Å resolution. The normal forces ($F_n$) between the two surfaces are measured by monitoring the bending of normal springs (normal spring constant $K_n$ = 210 N/m) via the changes in D arising from the applied motion. The lateral forces ($F_s$) are determined by the bending of shear springs with spring constant $K_s$ = 354 N/m, whose bending is monitored by an air-gap capacitance probe. All of the measurements were carried out cross 2 mM $Ca(NO_3)_2$ dispersions of the different species (PTN, LUV, NLG) following 2 h incubation. All SFB data shown are based on independent experiments (different pairs of mica sheets). Force profiles are normalized as ($F_n/R$) vs. D, which in the Derjaguin approximation corresponds to the interaction-energy/unit area between flat parallel surfaces obeying the same force vs. distance law.

## S4. Determining the bilayer thickness in cryo-TEM images

Cryo-TEM was used to determine the thickness of the lipid bilayer. Briefly, high-quality cryo-TEM images of the vesicles were acquired, and individual bilayers were selected for analysis. A straight line was drawn perpendicular to the membrane plane across a well-resolved region of the bilayer, and the corresponding intensity profile was extracted using Image J software. The bilayer appears as two distinct peaks in the intensity profile (see figure S1), corresponding to the electron-dense headgroup regions of the inner and outer leaflets[3]. The lipid bilayer thickness was defined as the distance between the centers of these two peaks, using the calibrated pixel size (3.17 pixel/nm) of the microscope to convert from pixels to nanometers. Multiple profiles were collected from different vesicles and membrane regions to obtain an average bilayer thickness and associated standard deviation.

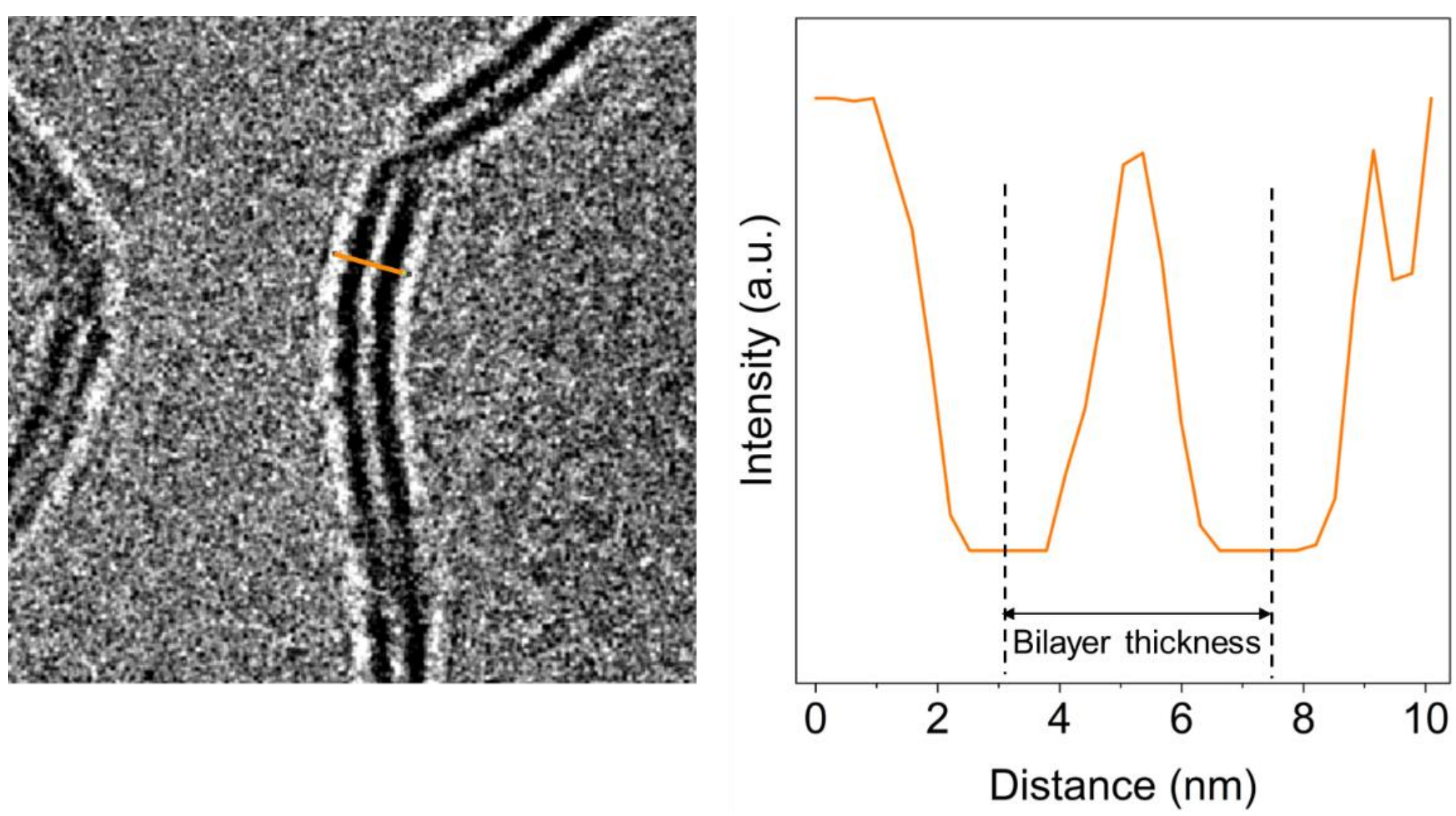

**Figure S3.** Representative cryo-TEM image (left) of LUV and corresponding intensity profile (right) taken along a line across the bilayer. The lipid bilayer thickness is defined as the distance between the centers of the two intensity peaks, as indicated by the dashed lines in the right panel. For each vesicle, we measured 3~5 positions, resulting in a total of 130 measurement points.

## S5. Self-healing property of lubrication

Traces a, b are the transmitted shear force traces, for NLGs at different compressions on second approach. The frequency analysis of traces show that the shear force slightly decreased over time.

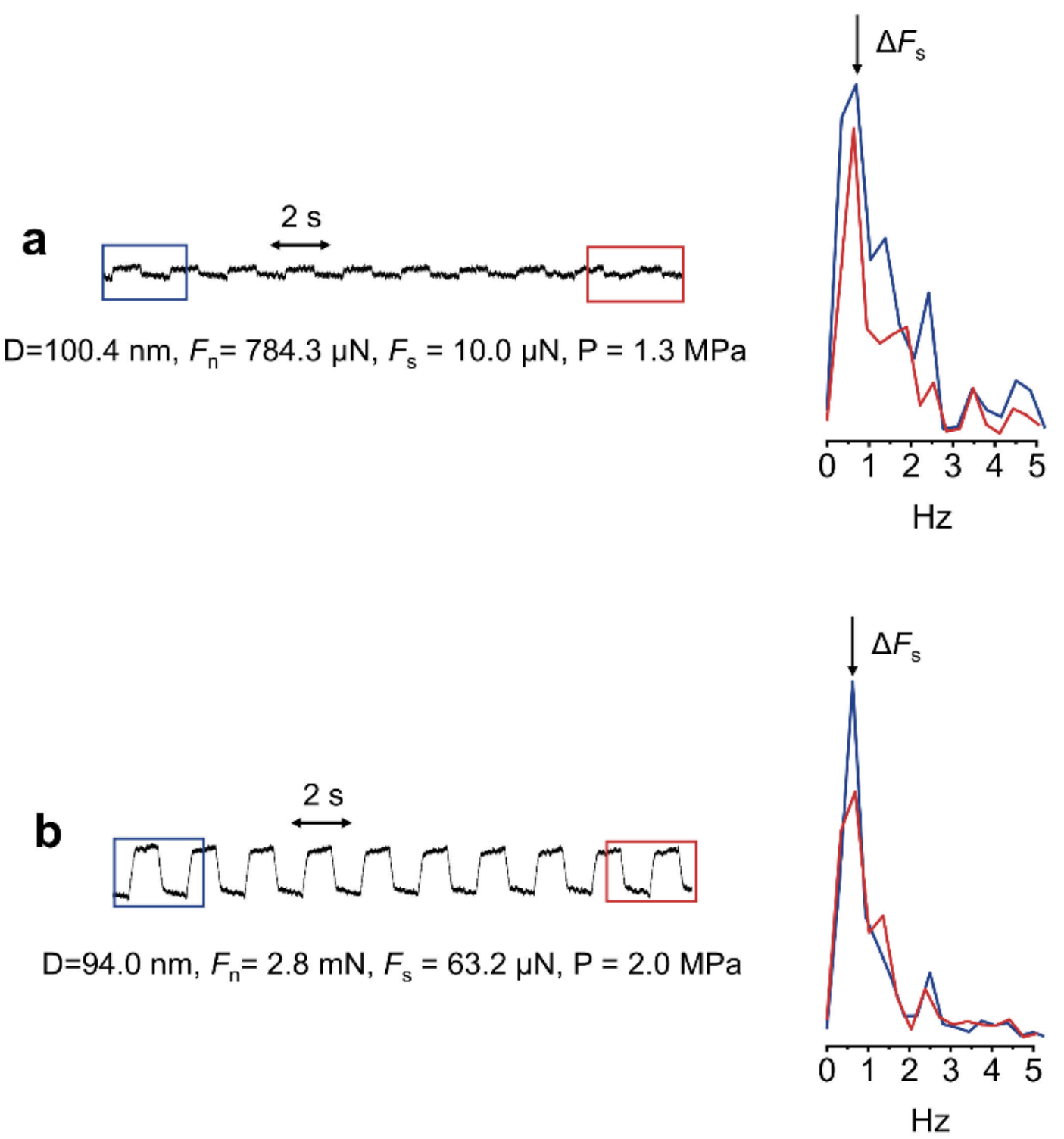

**Figure S4**. Self-healing of Lubrication of NLGs on the second approach. Shear traces (left) and their corresponding fast fourier transform (FFT, right) of NLGs under different pressures. The blue and red FFT traces correspond to the left and right-hand regions of the friction-force traces (blue and red rectangles), showing the reduction with sliding time of the friction force.

## S6. Velocity-dependent $F_s$

The velocity was controlled by changing frequency of the lateral back-and-forth motion (upper zig-zag in fig. 6A) while maintaining a constant amplitude $\Delta x_0$. The round (blue) symbols were collected at D ≈ 107 nm and P ≈ 2 MPa, the square (black) symbols show data obtained at D ≈ 93 nm and P ≈ 2.3 MPa, respectively a little below and a little above the irreversible transition described in the text.

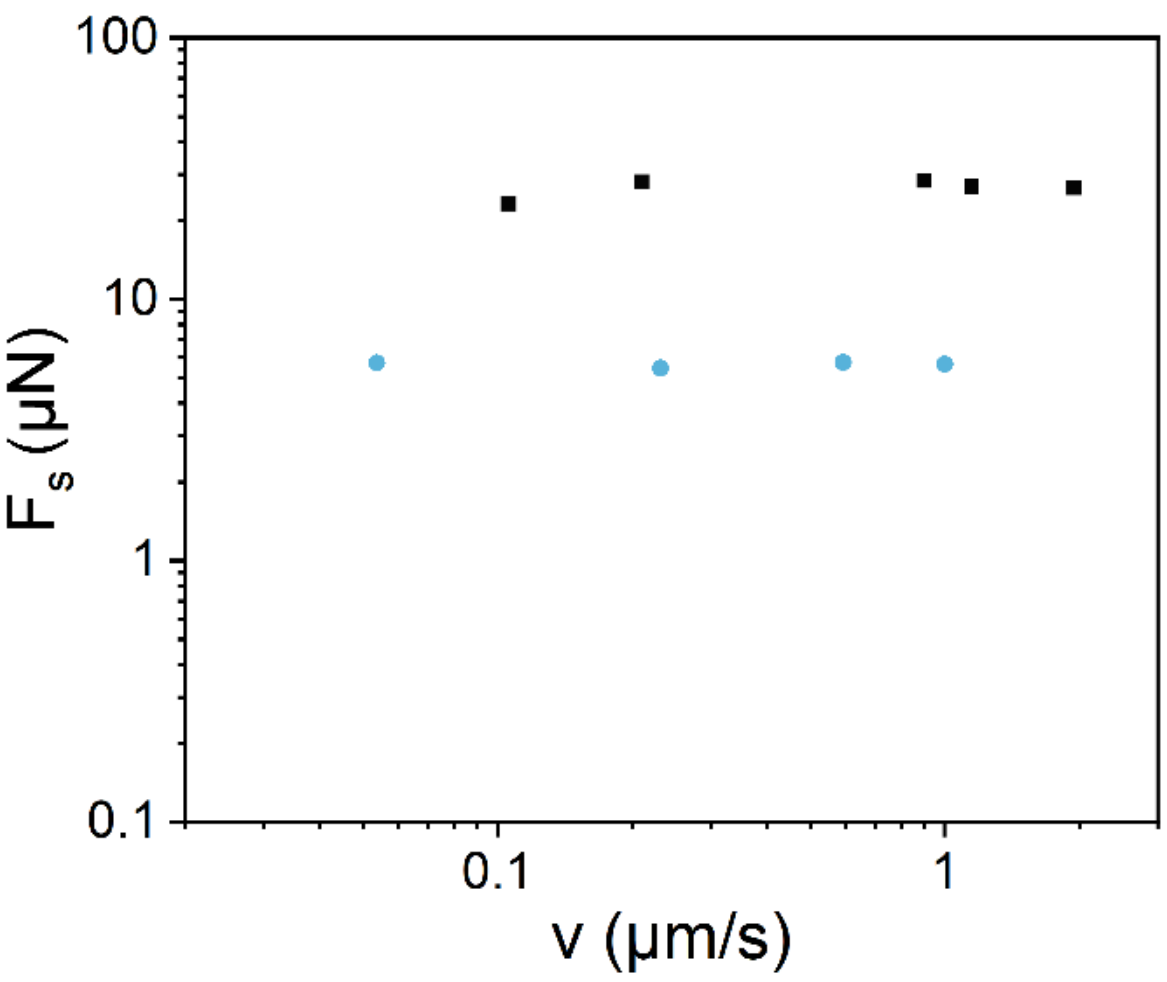

**Figure S5**. Representative shear forces ($F_s$) versus sliding velocity ($v$) of surfaces across NLGs.

## S7. Mechanical stabilization of NLG by encapsulated nanogels

Unfunctionalized HSPC vesicles (i.e. unPEGYlated LUVs) adsorb on mica from aqueous medium in an intact but metastable state (Figure S5). This is readily seen by considering the balance of interfacial energies. For an adsorbed, assumed-largely-flattened vesicle, i.e. where the adsorbed thickness is much less than the unperturbed vesicle, the total energy is the sum of interfacial plus distortion energy of the vesicle. The former may be approximated as ($A\gamma_{lipid/mica} + 3A\gamma_{lipid/water}$ ) where A is the contact area and $\gamma_{a/b}$ (< 0) is the interaction energy/unit area between *a* and *b*. The distortion energy for a highly flattened but intact vesicle is the bending energy integrated around the highly curved vesicle edge, and equals $-A(\gamma_{lipid/mica} - \gamma_{lipid/water})$, and is released once the vesicle ruptures. Clearly $|\gamma_{lipid/mica}| > |\gamma_{lipid/water}|$ for adsorption (and vesicle distortion) to occur in the first place, so that after rupture to a single bilayer of area ~2A equal to that of the membrane of the original vesicle, the total interfacial energy becomes $2(A\gamma_{\text{lipid/mica}} + A\gamma_{\text{lipid/water}}) < (A\gamma_{lipid/mica} + 3A\gamma_{lipid/water})$, the original interfacial energy. Thus rupture is energetically favourable and the adsorbed, non-PEGylated vesicle is therefore in a metastable state when intact.

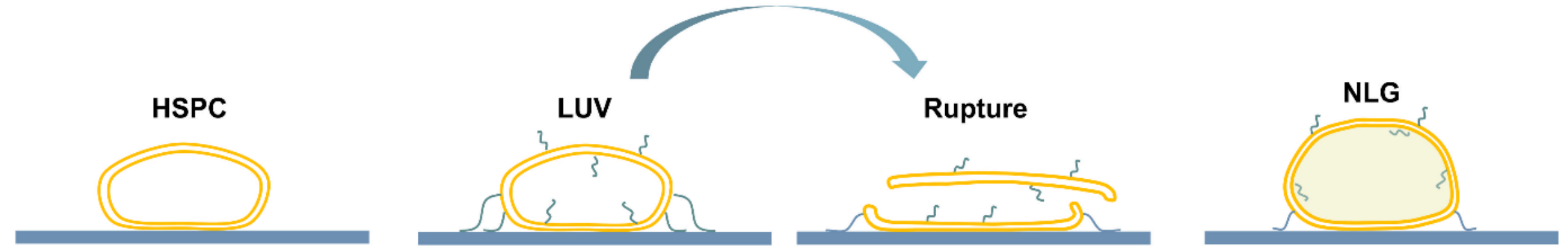

**Figure S6**. Schematic showing proposed rupture process of PEGylated LUVs on mica contrasted with stable NLGs.

Once vesicles are PEGylated (to ensure their colloidal stability in dispersion via steric repulsion), then on adsorption to a negative surface from the 2 mM $Ca^{++}$ salt solution the PEG moieties adsorb as well, as indicated in panel D of fig. 3 in the main text. This adsorption is promoted by chelation of the PEG's slightly-negatively-charged ether oxygens by $Ca^{++}$ ions which also bind to negatively charged surface sites, effectively "bridging" the PEG polymer to the surface. This bridging significantly enhances the PEG adsorption relative to adsorption from salt-free water, resulting in tension on the PEGylated DSPE lipids; in addition the negatively-charged DSPE headgroups also undergo a strong $Ca^{++}$-mediated bridging attraction to the mica. In consequence, the PEG-terminated DSPE lipids in the vicinity of the adsorbing mica are extracted from the vesicle bilayer membrane, disrupting the bilayer and leading to its rupture to the lower-free-energy single-bilayer ruptured state as discussed above. This is supported by a more detailed consideration of the rupture kinetics, as in ref.[4]

For the case where the vesicle encapsulates nanogels, as for the NLGs, the situation is fundamentally different: the vesicle membrane is now strongly attached to the enclosed nanogels via hydrogen bonding and cation-π interactions, in analogy to the cytoskeleton anchoring its membrane. This not only suppresses significant distortion of the NLG on adsorption, due to the additional energy required for nanogel compression, but in particular prevents rupture of the bilayer membrane even if PEG-terminated DSPE lipids in the vicinity of the adsorbing surface are extracted from the bilayer membrane. This is because the membrane is extensively ‘stitched’ to the enclosed nanogel and is therefore in a stable rather than metastable state when adsorbed, so that point disruptions due to removal of the PEG-terminated DSPE lipids would not result in overall membrane rupture.

## S8. $Ca^{2+}$ uptake by PTN

We diluted 1 mL PTN solution into 19 mL 2 mM $Ca^{2+}$ solution, so the concentration (mole) of $Ca^{2+}$ ions is:

$$n_{\text{Ca, total}} = 0.019 \text{ L} \times 2\times10^{-3} \text{ mol/L} = 3.8\times10^{-5} \text{ mol}$$

The total OH sites of TA in 1 ml PTN dispersion is:

$$n_{OH,TA} = f_{TA} \times \frac{m_{TA}}{M_{TA}} = 25 \times \frac{3\times10^{-3}}{1701} \approx 4.4 \times 10^{-5} \text{ mol}$$

where $f_{\text{TA}}$ is the number of phenolic OH per TA molecule, $m_{\text{TA}}$ is the mass of TA in 1mL PTN solution, and $M_{\text{TA}}$ is the molecular weight of TA.

As noted in section 4, only a fraction $\varphi$ (0.15) actually forms PVA-TA bonds, so the remaining OH sites from TA is:

$$n_{remaining\ OH,TA} = (1-\varphi)n_{OH,TA} \approx 3.7 \times 10^{-5} \text{ mol}$$

The total OH sites from PVA in 1 ml PTN dispersion is:

$$n_{OH,PVA} = \alpha \times \frac{m_{PVA}}{M_{PVA}} = 0.99 \times \frac{1\times10^{-2}}{44} \approx 2.3 \times 10^{-4} \text{ mol}$$

where α is the degree of hydrolysis of PVA, $m_{\text{PVA}}$ is the mass of PVA in 1mL PTN solution, and $M_{\text{PVA}}$ is the molecular weight of PVA.

The concentration of remaining OH sites from PVA is:

$$n_{remaining\ OH,PVA} = (1-\varphi)n_{OH,PVA} \approx 1.9 \times 10^{-4} \text{ mol}$$

The total OH sites which could then bridge with $Ca^{2+}$ in 1 mL PTN dispersion is:

$$n_{OH,total} = n_{remaining\ OH,PVA} + n_{remaining\ OH,TA} \approx 2.3 \times 10^{-4} \text{ mol}$$

If each $Ca^{2+}$ effectively binds ~2 OH groups, then the binding-relevant ratio is:

$$\frac{n_{OH,total}}{n_{\text{Ca, total}}} \approx 3$$

Thus, PTN contains approximately three times more OH groups than required to coordinate all $Ca^{2+}$ ions, assuming a 2 OH:1 $Ca^{2+}$ binding stoichiometry. In this case, the zeta potential of the PTNs remained negative after aggregation, hindering their adsorption onto the negatively charged mica surface (Figure S6).

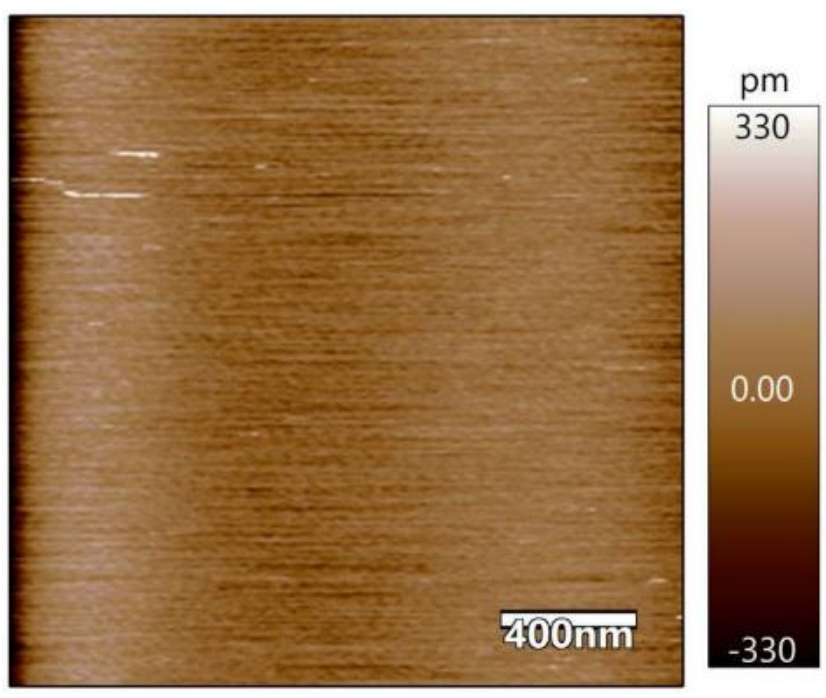

**Figure S7.** Height AFM image of mica incubated in PTNs in 2 mM $Ca(NO_3)_2$ solution.

## S9. Double-layer electrostatic interactions and Debye screening length

Derjaguin–Landau–Verwey–Overbeek (DLVO) theory[5] derives for the normalized force between curved charged surfaces (mean radius of curvature *R*) in aqueous salt solution at closest surface separation *D* the following expression:

$$\frac{F(D)}{R} = 64Ck_BT\kappa^{-1}tanh^2\left(\frac{e\psi_0}{k_BT}\right)exp(-kD) - \frac{A_H}{12\pi D^2}$$

Here *C* is the effective electrolyte concentration in mol/dm$^3$, *T* is the temperature (296 ± 1 K), $k_B$ is Boltzmann's constant, $A_H$ is the Hamaker constant of mica across water ($2 \times 10^{-20}$ J), $\psi_0$ (78 mV) is the effective (large-separation) surface potential and $\kappa^{-1}$ is the Debye screening length corresponding to the salt concentration used. For a divalent ion such as $Ca^{++}$ at room temperature (25C), the expression for the screening length in nm is $\kappa^{-1} = 0.304/(3C)^{0.5} \approx 3.9$ nm for the 2 mM calcium salt concentration used in our experiments.

## S10. H-bond energy density

Energy density, $u_{\mathrm{HB}}$ , is the enthalpy stored in hydrogen bonds per volume of hydrogel:

$$u_{\mathrm{HB}} = n_{\mathrm{HB}}^{\mathrm{V}} \times \Delta H_{\mathrm{HB}}$$

where $n_{\mathrm{HB}}^{\mathrm{V}}$ is the moles of formed PVA-TA H-bonds per unit volume. $\Delta H_{\mathrm{HB}}$ is the enthalpy per H-bond (phenol-alcohol H-bonds are commonly ~12-25 kJ mol$^{-1}$, here we use 18 kJ mol$^{-1}$).

The OH sites (per 100 g gel) from PVA (the degree of hydrolysis α is 0.99) are:

$$n_{OH,PVA} = \alpha \times \frac{m_{PVA}}{M_{PVA}} = 0.99 \times \frac{19.2}{44} \approx 0.432 \text{ mol}$$

where $m_{\mathrm{PVA}}$ is the mass of PVA in 100 g of gel, and $M_{\mathrm{PVA}}$ is the molecular weight of PVA.

The OH sites (per 100 g gel) from TA ($f_{\mathrm{TA}}$ is the number of phenolic OH per TA molecule, commonly taken ~25) are:

$$n_{OH,TA} = f_{TA} \times \frac{m_{TA}}{M_{TA}} = 25 \times \frac{5.8}{1701} \approx 0.085 \text{ mol}$$

where $m_{\mathrm{TA}}$ is the mass of TA in 100 g of gel, and $M_{\mathrm{TA}}$ is the molecular weight of TA.

The maximum number of PVA-TA pairs is limited by the scarcer OH pool:

$$n_{pairs,max} = \min\,(n_{OH,TA},\ n_{OH,PVA}) = 0.085 \text{ mol}$$

Only a fraction $\varphi$ of the total -OH groups actually forms PVA–TA bonds; the rest are free or H-bonded to water/PVA–PVA/TA–TA. For highly hydrated gels (≈70-90 wt% water) we expect $\varphi \approx 0.15$-$0.35$. For the nanogel in our system, we expect more OH groups to be exposed on the outside, so that $\varphi \approx 0.15$. Then:

$$n_{HB} = \varphi \times n_{pairs,max} = 0.0128 \text{ mol}$$

Per-volume bond density:

$$n_{\mathrm{HB}}^{\mathrm{V}} = \frac{n_{HB}}{V} = \frac{n_{HB}\rho_{gel}}{m_{gel}} \approx 128 \text{ mol m}^{-3},\ \rho_{gel} \approx 1.0 \text{ g mL}^{-1}$$

H-bond energy density of nanogel:

$$u_{\mathrm{HB}} = 128 \times 18 \approx 2.3 \text{ MJ m}^{-3}$$

## S11. Confocal colocalization analysis

Dual-labeled NLGs were imaged by confocal microscopy, with the lipid bilayer labeled by DiD and the interior nanogel labeled by RhB. Raw two-channel confocal images were opened in Fiji/ImageJ and split into individual grayscale channels. The separated channels were converted to 32-bit format before background correction to avoid intensity wraparound artifacts. Background correction was applied identically to both channels, and negative pixel values generated during background subtraction were set to zero. The same region of interest (ROI) was applied to both channels using the Fiji ROI Manager. Pixel-wise correlation between the DiD and RhB fluorescence channels was quantified within the selected ROI using Fiji Coloc 2. Pearson's correlation coefficient (PCC) is one of the standard procedures in component recognition for matching one component with another and can be used to describe the degree of overlap between two components. It provides information about the similarity of shape without regard to the average intensity of the signals.[6] The formula for PCC is given below for a typical image consisting of two components:[6, 7]

$$PCC = \frac{\sum_i (R_i - \bar{R}) \times (G_i - \bar{G})}{\sqrt{\sum_i (R_i - \bar{R})^2 \times \sum_i (G_i - \bar{G})^2}}$$

where $R_i$ and $G_i$ refer to the intensity values of RhB and DiD, respectively, of pixel $i$, and $\bar{R}$ and $\bar{G}$ refer to the mean intensities of RhB and DiD, respectively, across the ROI. PCC values range from 1 for two components whose fluorescence intensities are perfectly, linearly related, to -1 for two components whose fluorescence intensities are perfectly, but inversely, related to one another.[7] Values near zero reflect distributions of probes that are uncorrelated with one another. The paired pixel intensities were visualized as a scatter plot (Figure S8), which showed a positively associated fluorescence population. This indicates that regions with higher lipid-associated DiD signal generally coincided with increased RhB signal from the interior nanogel.

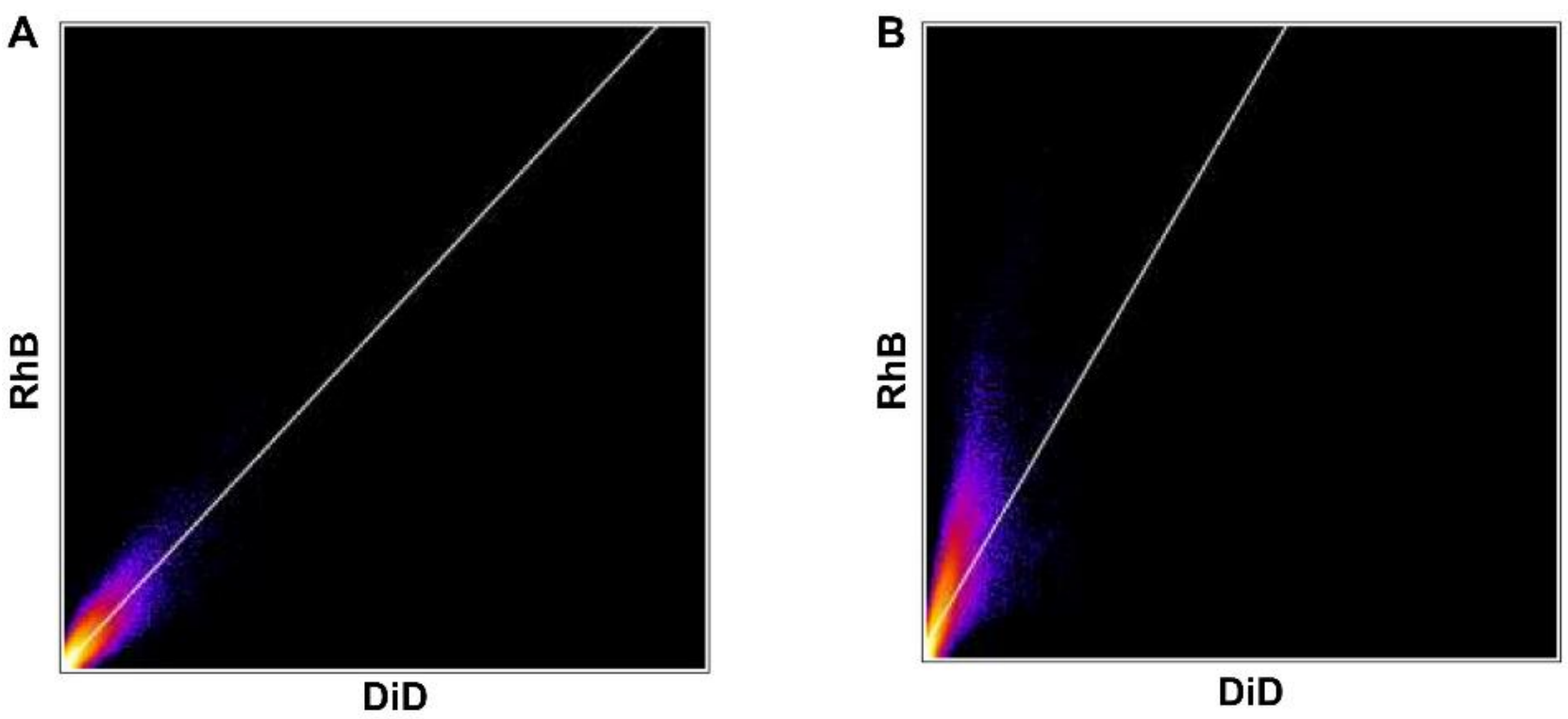

**Figure S8**. Scatterplot of DiD and RhB pixel intensities of reresentive ROI in the non-wear (A) and wear (B) region. The intensity of a given pixel in the DiD image is used as the x-coordinate of the scatter plot and the intensity of the corresponding pixel in the RhB image as the y-coordinate.

Line-profile analysis was performed in Fiji/ImageJ to compare the local spatial distributions of the DiD-labeled lipid bilayer and RhB-labeled nanogel. A straight line ROI was drawn across representative NLG-covered regions. Fluorescence intensity values along the line were extracted separately from the DiD and RhB channels using the "Plot Profile" function. The extracted intensity profiles were exported and normalized to their respective maximum values to compare the spatial positions of fluorescence peaks independent of differences in absolute signal intensity:

$$I_{norm} = \frac{I - I_{min}}{I_{max} - I_{min}}$$

where $I$ is the measured fluorescence intensity along the line, and $I_{\min}$ and $I_{\max}$ are the minimum and maximum intensities of that channel along the same line. The normalized RhB and DiD intensity profiles were plotted as a function of distance along the line. PCC was calculated to evaluate the relationship between the fluorescence intensities of RhD and DiD (Figure 7E).

The relative fluorescence retention of the lipid and nanogel components was evaluated from confocal line-profile data. DiD and RhB fluorescence intensity profiles were extracted from the same line ROIs in non-wear and wear regions using Fiji/ImageJ, with three independent line profiles analyzed for each region. The DiD/RhB intensity ratio ($I_{DiD/RhB}$) was calculated from raw, background-corrected fluorescence intensities:

$$I_{DiD/RhB} = \frac{\sum I_{DiD}}{\sum I_{RhB}}$$

To avoid artifacts from near-zero denominator values, only signal-positive positions were included, defined as positions where both DiD and RhB intensities were greater than 10% of their respective maximum intensity along the same line profile. The pointwise ratio $I_{DiD}/I_{RhB}$ was calculated at each signal-positive position, and the median ratio for each line was used for comparison. Regional values were reported as mean ± SD from three line profiles. The $I_{DiD}/I_{RhB}$ was interpreted as a relative fluorescence intensity ratio reflecting lipid-associated signal relative to nanogel-associated signal, not as an absolute compositional ratio.

## S12. *In vitro* cytotoxicity

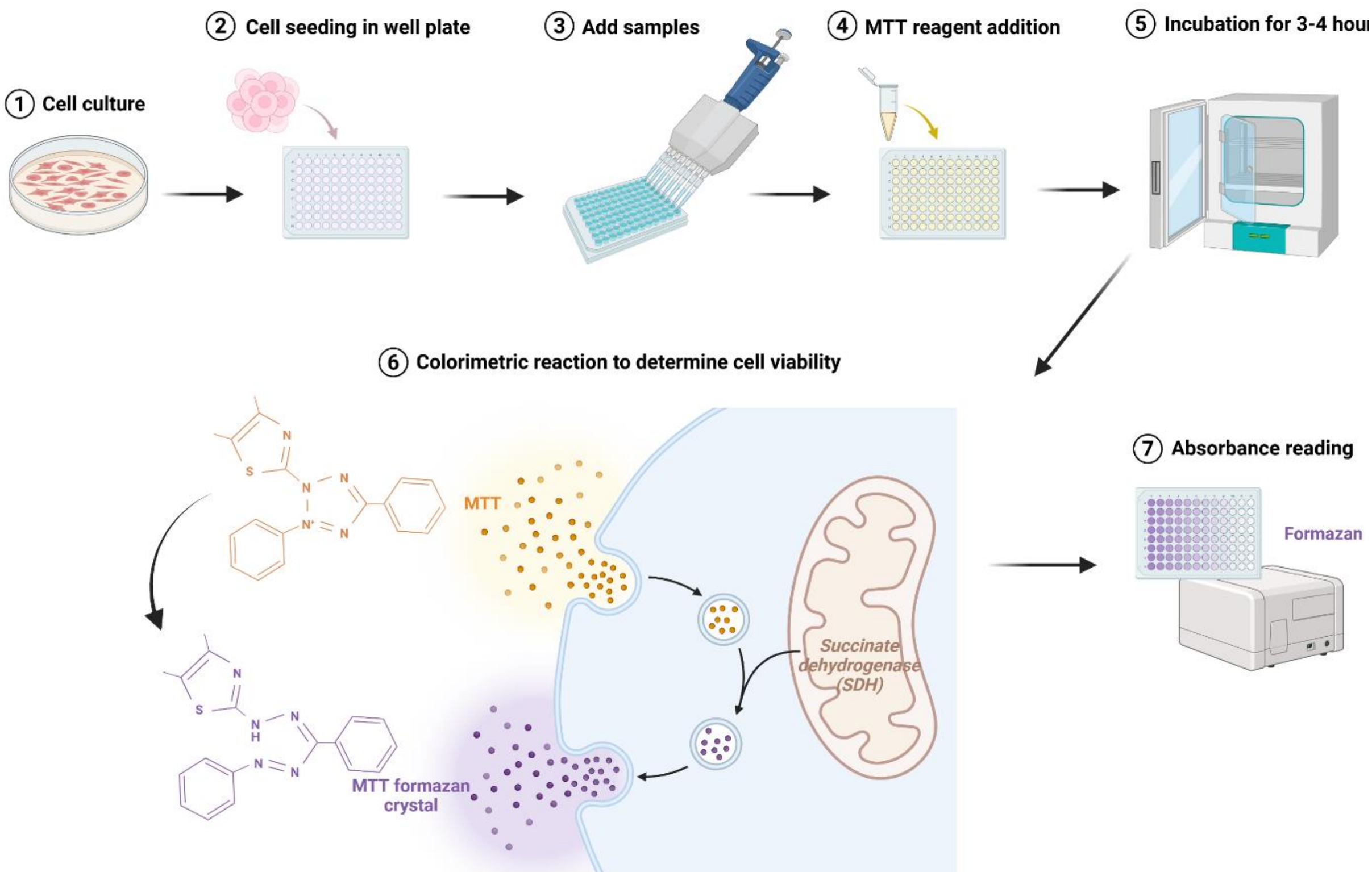

**Figure S9**. Schematic representing the procedure for *in vitro* cytotoxicity evaluation (figure was created by BioRender).

## S13. Drug loading

RhB standard solutions were prepared in pure water at concentrations of 1, 2, 4, 6, 8, and 10 μg/mL. Then the absorbance values of the solutions were scanned within the range of 200−800 nm of the absorption spectrum using a UV−vis spectrophotometer. The equation of the linear regression curve obtained at ~560 nm was y = 0.185x - 0.017, where y is the absorbance of the sample solution and x is the concentration of RhB present in the sample (Figure S11). RhB showed a linear response with a correlation coefficient ($R^2$) of 0.999. The amount of non-trapped RhB was quantified from the collected supernatant and washing solutions using the calibration curve. The RhB loading in the NLGs was calculated by subtracting the amount of RhB detected in the supernatant and washing solutions from the initial amount of RhB used during NLG preparation.

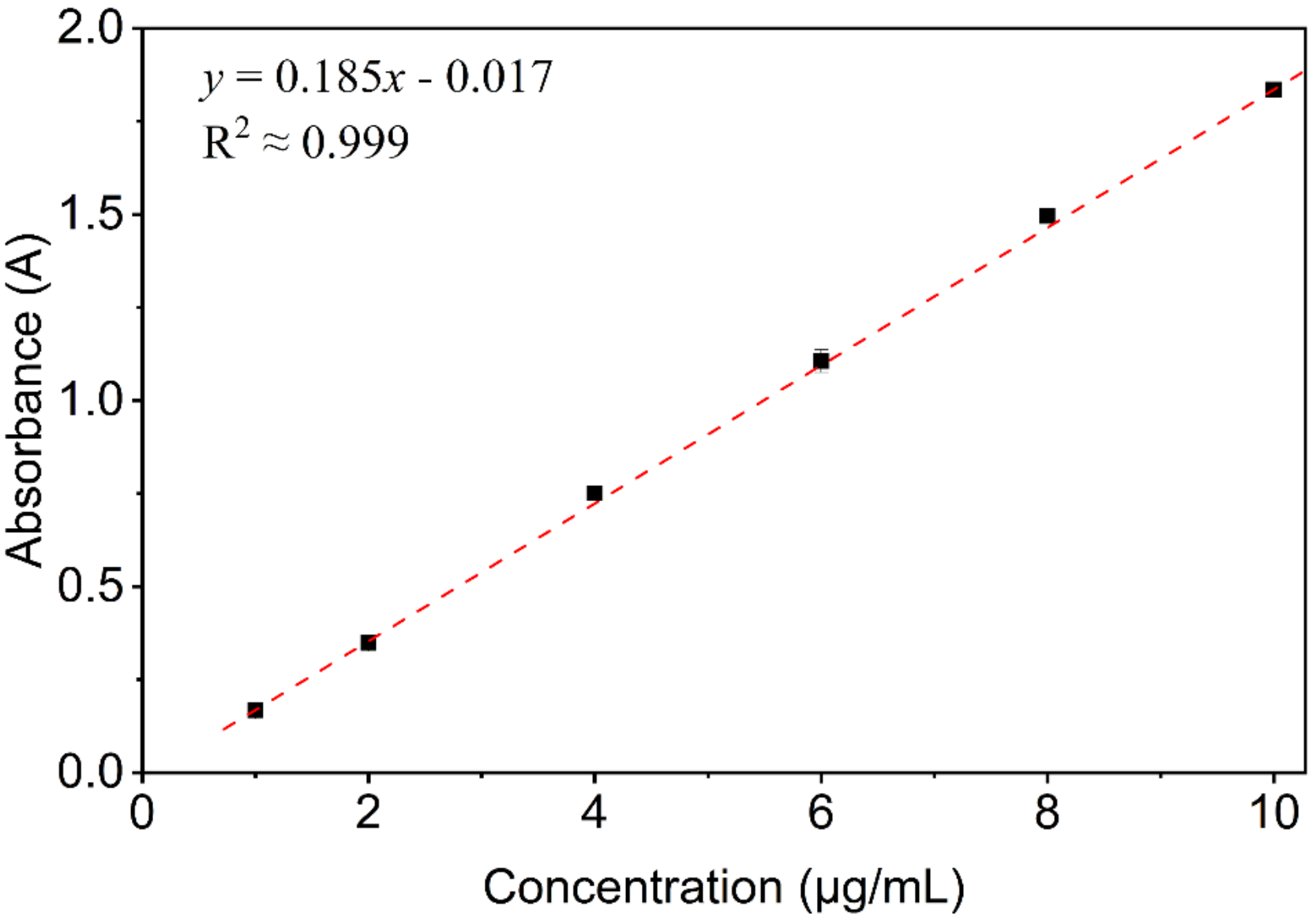

**Figure S10**. Absorbance *vs.* concentration calibration curve of RhB. Data are means of three independent assays. The regression equation and correlation coefficient are also shown.

**Table S1.** DLS results of PTN, LUV and NLG in pure water.

| Samples | Peak value (nm) | PDI |
|---|---|---|
| PTN | 332.1±6.3 | 0.227±0.011 |
| LUV | 134.3±2.1 | 0.087±0.028 |
| NLG | 151.8±3.1 | 0.197±0.008 |

**Table S2**. Zeta potential of LUV and NLG in 0.15 M $NaNO_3$.

| Samples | Zeta potential (mV) |
|---|---|
| LUV | −2.1 ± 1.2 |
| NLG | -1.9 ± 0.2 |